\documentclass[final,5p,twocolumn]{elsarticle}
\usepackage{amsmath,amssymb,amsfonts}
\usepackage{orcidlink}
\usepackage{graphicx}
\usepackage{booktabs}
\usepackage{hyperref}
\usepackage{cleveref}
\crefname{appsec}{}{}\Crefname{appsec}{}{}
\usepackage{listings}
\usepackage{xcolor}
\usepackage{placeins}
\usepackage{bm}
\usepackage{mathtools}
\usepackage{tikz}
\usetikzlibrary{arrows.meta,decorations.markings,positioning,fit,backgrounds,calc,shapes.geometric}
\usepackage{lineno}
\usepackage{enumitem}

\definecolor{codegreen}{rgb}{0,0.6,0}
\definecolor{codegray}{rgb}{0.5,0.5,0.5}
\definecolor{codepurple}{rgb}{0.58,0,0.82}
\definecolor{backcolour}{rgb}{0.97,0.97,0.97}

\lstdefinestyle{python}{
    backgroundcolor=\color{backcolour},
    commentstyle=\color{codegreen},
    keywordstyle=\color{codepurple},
    numberstyle=\tiny\color{codegray},
    stringstyle=\color{codegreen},
    basicstyle=\ttfamily\small,
    breaklines=true,
    numbers=left,
    numbersep=5pt,
    language=Python,
    frame=single,
    framerule=0.5pt,
}
\tikzset{
  Cprop/.style={thick, color=blue!70!black},
  Rprop/.style={thick, color=red!70!black, dashed,
    decoration={markings, mark=at position 0.55 with {\arrow{Stealth[length=5pt]}}},
    postaction={decorate}},
  extnode/.style={circle, fill=black, inner sep=2pt, minimum size=5pt},
  vertnode/.style={rectangle, fill=black, inner sep=2.5pt, minimum size=6pt},
  nlabel/.style={font=\small},
}

\newcommand{\sftwick}{\textsc{sft-wick}}
\newcommand{\vev}[1]{\langle #1 \rangle}
\newcommand{\vevfree}[1]{\langle #1 \rangle_{S_0}}
\newcommand{\Sint}{S_{\mathrm{int}}}
\newcommand{\dd}{\mathrm{d}}
\newcommand{\ii}{\mathrm{i}}
\newcommand{\ito}{It\^{o}}
\newcommand{\DeterOp}{\hat{\mathcal{D}}}
\newcommand{\RespOp}{\mathcal{R}}
\newcommand{\CorrOp}{\mathcal{C}}

\newcommand{\cumulant}{\mathcal{K}}

\newenvironment{proof}[1][Proof]{\par\noindent\textit{#1.}\ }{\hfill$\square$\par\medskip}

\journal{Computer Physics Communications}

\begin{document}

\begin{frontmatter}

\title{\sftwick: A formalism and package for Feynman-diagram expansion and 
evaluation in stochastic field theories}

\author[jbca]{Zheng Zhang\,\orcidlink{0000-0002-9154-2803}}
\ead{zheng.zhang@manchester.ac.uk}

\address[jbca]{Jodrell Bank Centre for Astrophysics, University of Manchester,
  Manchester, M13 9PL, United Kingdom}

\begin{abstract}
When stochastic field dynamics are cast into a path-integral formulation, 
perturbation theory becomes systematic but the resulting expansion quickly 
grows combinatorially large. 
The setting targeted here includes fields with several components
and several dimensions, matrix propagators,
tensor-valued couplings, and non-Gaussian driving noise specified by
arbitrary $n$-point cumulants. Wick pairings grow factorially, and
component indices must be routed through the tensor-valued vertices.
The useful output is not a raw contraction list, but a diagram table:
one entry per topology, with multiplicities, coupling sums, signs, and
causal constraints resolved.
We present \sftwick{}, an open-source Python package that constructs
these diagram tables and computes their integrals numerically. Given
an action and an observable, it enumerates topologically distinct
Feynman diagrams, derives their algebraic coefficients, and evaluates
the resulting diagram integrals from user-supplied response and
cumulant functions.
The core algorithm enumerates spatial topologies before routing
component indices, avoiding contraction-by-contraction Wick expansion.
Response-field constraints, including vanishing response-response
contractions, the \ito{} prescription, and the absence of causal
response loops, are enforced during enumeration. Predictions are
validated against direct Langevin simulations in two worked examples,
with Gaussian and with non-Gaussian driving noise, and agree with the
simulations within stated error budgets.
\end{abstract}

\begin{keyword}
Stochastic Differential Equations \sep
Martin--Siggia--Rose Formalism \sep
Feynman Diagrams \sep
Perturbation Theory \sep
Symbolic Computation
\end{keyword}

\end{frontmatter}

\section{Introduction}
\label{sec:introduction}

The Martin--Siggia--Rose (MSR) formalism
\cite{Martin1973,Janssen1976,DeDominicis1976} recasts stochastic
differential equations into the language of field theory. At its core,
it provides a systematic perturbative machinery for determining how
the statistical properties of stochastic driving fields are imprinted,
through nonlinear dynamics, onto the observable fields of interest.
The formalism was born from the turbulence closure problem: the nonlinear
convective term in the Navier--Stokes equations couples different scales of
motion, creating an infinite hierarchy in which the equation for the
$n$-point correlation function involves the $(n{+}1)$-point function.
Early diagrammatic field-theoretic treatments of this hierarchy
\cite{Wyld1961} were placed on a systematic footing when Martin, Siggia,
and Rose, by introducing an auxiliary response field, organised the
hierarchy into a self-consistent diagrammatic perturbation theory
\cite{Martin1973}, later reformulated independently by
Janssen \cite{Janssen1976} and De~Dominicis \cite{DeDominicis1976}. This
``MSR--Janssen--De~Dominicis'' framework underpins the field-theoretic
renormalisation-group treatment of critical dynamics
\cite{DeDominicis1978} and continues to be used in field-theoretic
approaches to turbulence \cite{Forster1977}.

Since its origins in fluid dynamics, the MSR formalism has been applied
across a range of areas of physics. In \emph{stochastic
inflation}, the coarse-graining procedure
\cite{starobinsky2005stochastic} maps sub-horizon quantum fluctuations of the
inflaton to classical noise driving the long-wavelength modes,
enabling non-perturbative computation of the statistics of curvature
perturbations \cite{Vennin2015}. In \emph{cosmological structure
formation}, Kinetic Field Theory (KFT) uses a closely related
response-field path integral over microscopic $N$-body phase-space
trajectories, yielding a description of density fluctuations that
remains well-defined beyond shell-crossing \cite{Bartelmann2016}.
In \emph{critical dynamics} and \emph{reaction--diffusion systems}
\cite{Tauber2014}, the MSR action exhibits a hidden (BRST) supersymmetry
in equilibrium Langevin dynamics connected to the
fluctuation--dissipation theorem \cite{Hertz2017}, with
dimensional-reduction antecedents in random-field problems
\cite{Parisi1979}. The formalism has also found
applications in \emph{neuroscience}, where networks of coupled
stochastic differential equations describe neuronal dynamics
\cite{ChowBuice2015}. At the quantum--classical interface, the MSR auxiliary field
is the classical limit of the ``quantum'' component of the Keldysh
(Schwinger--Keldysh) formalism \cite{Kamenev2023,Altland2010}; we
return to this correspondence in \cref{sec:propagators}.

Despite this breadth, existing pedagogical treatments of the MSR
path integral tend to specialise: applied introductions work with
single SDEs or a few scalar fields \cite{ChowBuice2015,Hertz2017},
while multi-component textbook treatments sit inside the
Hohenberg--Halperin critical-dynamics classification with Gaussian
driving noise \cite{Tauber2014}. The weak lensing problem that
motivated this work \cite{Zhang2025_SFTLensing} requires the
combination: stochastic partial differential equations (SPDEs) for
multi-component fields with tensor-valued couplings, matrix
propagators carrying non-trivial component-index structure, and
non-Gaussian driving statistics entering through higher-order
cumulants. Each ingredient has appeared individually in the MSR
literature, but to our knowledge they have not previously been
assembled into one explicit prescription. The present work therefore
serves a dual purpose: it gives a self-contained, consolidated
formulation of this multi-component, non-Gaussian MSR formalism in
the operator-bookkeeping form the software consumes
(\cref{sec:theory}), and it introduces
\sftwick{}\footnote{\url{https://github.com/StatFieldTheory/sft-wick}}
to automate the resulting perturbative calculations.

In the MSR formalism, the statistical average of an observable $O$ with
respect to an action $S = S_0 + \Sint$ is computed via the perturbative
expansion
\begin{equation}
  \label{eq:perturbative-expansion}
  \vev{O}_S
  = \sum_{n=0}^{N} \frac{(-1)^n}{n!}\,
  \vevfree{O\,\Sint^n}\,,
\end{equation}
where each term is evaluated using Wick's theorem applied to the free
(Gaussian) theory $S_0$. The field content comprises physical fields
$\varphi$ and conjugate response fields $\psi$, with the crucial MSR
constraint that response-response contractions vanish:
$\vev{\psi\,\psi}_{S_0} = 0$. Only the correlation propagator
$C = \vev{\varphi\,\varphi}_{S_0}$ and the retarded response propagator
$R = \vev{\varphi\,\psi}_{S_0}$ survive.

While the formalism is elegant, practical calculations become rapidly
intractable at higher perturbative orders. The number of complete
Wick pairings of $2n$ field operators scales as
$(2n-1)!! = 1 \cdot 3 \cdot 5 \cdots (2n-1)$, which grows
super-exponentially and exceeds $10^5$ already at $n=7$. For
multi-component theories, each pairing further branches into
distinct routings of the component indices through the coupling
tensors, compounding the cost.
Most of these pairings yield topologically redundant Feynman
diagrams. The key to tractability is to group contributions by
topology, so that many equivalent pairings are represented by a
single diagram with a known multiplicity, reducing the
combinatorial explosion to an enumeration of distinct graph
topologies.

Several symbolic computation tools are used for Feynman-diagram and
field-theory calculations in quantum field theory, notably
\textsc{FeynArts} \cite{Hahn2001} for diagram generation, together with
\textsc{Cadabra} \cite{Peeters2007}, and
general-purpose systems such as \textsc{SymPy} \cite{Meurer2017} for the
attendant symbolic manipulation. However, none of these tools natively
handles the constraints specific to the MSR formalism:
the vanishing of response-response correlators,
the \ito{} calculus prescription,
the elimination of causal loops in the retarded propagator graph,
or the two-propagator structure that distinguishes the correlation
propagator~$C$ from the response propagator~$R$.
Adapting general-purpose QFT tools to the MSR setting therefore requires
substantial manual intervention at each perturbative order.

The development of \sftwick{} was originally motivated by a companion
paper \cite{Zhang2025_SFTLensing}, in which we formulate gravitational
weak lensing as a statistical field theory within the MSR framework.
That application required the path-integral treatment to be generalised to 
the multi-component SPDE setting, involving $N$-component vector fields, 
tensor-valued coupling constants $F_{abc}$ and matrix propagators $C_{ab}$ 
and $R_{ab}$. This made an automated tool essential.

Although built for gravitational weak lensing, we recognised that the
resulting machinery is applicable to a much broader range of physical
problems. The package handles MSR-type stochastic field theories whose
interaction vertices are monomials in the fields contracted with
coupling tensors. Both local vertices (all fields at the same spatial
point) and non-local vertices (fields at different points with
kernel-valued couplings) are supported. Applications involving
derivative couplings can be accommodated by working in Fourier space,
where spatial derivatives become momentum-dependent vertex factors that
fit the monomial structure. The multi-component generalisation is
relevant whenever the underlying fields carry internal degrees of
freedom (e.g.\ spin, polarisation, or species labels). These cover a
broad range of scenarios, from critical dynamics and reaction--diffusion
systems \cite{Tauber2014} to cosmological perturbation theory
\cite{Bernardeau2002}. We therefore present \sftwick{} as a broadly
applicable package, decoupled from the specific cosmological problem
that prompted its development. The weak lensing formalism and physical
results are described in \cite{Zhang2025_SFTLensing}.

In this paper, we present \sftwick{}, an open-source Python package that
automates the complete perturbative calculation pipeline within the MSR
formalism. Given an action (field content and interaction vertices) and an
observable, \sftwick{} delivers two concrete outputs:
\begin{enumerate}
  \item \textbf{Diagram tables.} At each perturbative order, \sftwick{}
    enumerates all topologically distinct Feynman diagrams, groups
    equivalent Wick contractions by topology, and derives the algebraic
    coefficient for each diagram, including coupling-tensor sums,
    combinatorial prefactors, and response phase factors. The result is a
    structured list of \texttt{DiagramTerm} objects, each containing the
    propagator connectivity, summation indices, integration variables,
    and the fully simplified coefficient, ready for inspection or
    further processing.
  \item \textbf{Numerical predictions.} Given user-supplied propagator
    functions (the response function $R$ and the driving-field cumulant
    $\cumulant$), \sftwick{} evaluates each Feynman diagram numerically
    and sums the contributions to compute the observable at the desired
    perturbative order. The evaluation pipeline handles spatial
    structure decomposition, propagator caching, and multidimensional
    integration automatically.
\end{enumerate}

The central algorithmic contribution enabling output~(1) is a
\emph{hybrid contraction engine} that first enumerates topologically
distinct spatial structures with their combinatorial multiplicities,
and then generates the distinct component-index routings within each
topology. This avoids the full $(2n-1)!!$ enumeration of Wick
pairings and, together with automatic diagram isomorphism detection,
enables efficient computation at perturbative orders that would be
impractical by hand or with general-purpose QFT diagrammatic tools.

\paragraph{Pedagogical scope}
The formalism surveyed in \cref{sec:theory} is not a new theoretical
framework: the individual ingredients (the MSR path integral,
multi-component Langevin dynamics, and cumulant expansions for
non-Gaussian driving fields) are all established in the literature.
Their combination into a single systematic prescription for
multi-component SPDEs with arbitrary driving-field statistics is,
however, scattered across sources focused on scalar or specialised
problems.  To make \sftwick{} usable without tracing this literature,
we provide a self-contained derivation of the formalism in the form
actually consumed by the software: explicit operator bookkeeping,
conventions for the \ito{} prescription, and the reduction of
perturbative averages to a finite set of Feynman-diagram integrals.
Readers already familiar with MSR may skim \cref{sec:theory} and
proceed directly to the worked examples in
\cref{sec:examples,sec:example2} or the algorithmic description in
\cref{sec:algorithm}.

The remainder of this paper is organised as follows.
\Cref{sec:theory} reviews the theoretical background of the MSR
formalism and Wick's theorem.
\Cref{sec:examples} presents a worked two-component Langevin example
with Gaussian driving that validates \sftwick{} against direct numerical
simulation, giving the reader a concrete picture of the package's inputs
and outputs before the algorithmic details, and \cref{sec:example2}
extends it to a non-Gaussian driving field, whose cumulants of every
order enter as non-local vertices, with an explicit error budget for the
comparison with simulation and a free-field limit that is exact.
\Cref{sec:algorithm} then describes the algorithms and implementation
underlying these results, covering the contraction engines,
diagram collection, and the numerical evaluation pipeline.
\Cref{sec:summary} summarises and discusses future directions.

\section{Theoretical background}
\label{sec:theory}

This section provides a self-contained review of the MSR formalism to
establish notation and conventions used throughout the paper. Readers
familiar with the formalism may still wish to skim this section for how
the problem is framed: we present the MSR machinery as a generalisation
to the dynamics of multi-component fields with polynomial interactions,
showing how the cumulants of the driving field propagate, through the
perturbative expansion, to the moments and cumulants of the physical
fields.
This generalisation was developed for the weak lensing application
described in \cite{Zhang2025_SFTLensing}, but the framework implemented
in \sftwick{} is not specific to that problem.

\paragraph{Terminology}
Throughout this paper, we refer to the stochastic term $\eta$ in the
Langevin equation as the \emph{driving field}, to distinguish it from
the \emph{auxiliary currents} $J$ and $K$ that appear in generating
functionals as mathematical devices for computing correlation functions.
(We try to avoid the term ``source'' field, which can be ambiguous between the 
physical driving field and the auxiliary currents in generating functionals.) 

\subsection{MSR action for multi-component fields}
\label{sec:msr-action}

Consider a system of stochastic partial differential equations for
$N$-component fields $\varphi_a(\bm{x}, t)$, where $a = 1, \ldots, N$
labels field components, $t$ is an evolution parameter (typically time),
and $\bm{x}$ collectively denotes all remaining coordinates that index
the field degrees of freedom: for example, spatial position, Fourier mode, angular
coordinates on the sky, or any other continuous labels. When $\bm{x}$
is absent, the formalism reduces to a system of ordinary stochastic differential
equations for $\varphi_a(t)$.
The dynamics is governed by a
Langevin-type equation of the form
\begin{equation}
  \label{eq:langevin}
  \partial_t \varphi_a(\bm{x}, t) = A_{a b}(\bm{x}, t) \, \varphi_b(\bm{x}, t) 
  +  \mathcal{F}_a[\varphi] + \eta_a(\bm{x}, t)\,,
\end{equation}
where $A_{a b}$ encodes the linear dynamics (here and hereafter,
repeated Latin indices are summed over field components),
\begin{multline}
  \mathcal{F}_a[\varphi] = F^{\rm (3)}_{a b c}(\bm{x}, t)\,\varphi_b(\bm{x}, t)\,\varphi_c(\bm{x}, t) \\
  + F^{\rm (4)}_{a b c d}(\bm{x}, t)\,\varphi_b(\bm{x}, t)\,\varphi_c(\bm{x}, t)\,\varphi_d(\bm{x}, t) + \cdots
\end{multline}
collects the deterministic nonlinearities (polynomial in the fields),
and $\eta_a$ is a stochastic driving field with prescribed statistics.

The perturbative framework developed below assumes a zero-mean driving
field. When this is not the case, or when the lowest-order nonlinearity is 
quadratic or the initial conditions are non-trivial, 
a mean-field subtraction can be performed to bring the problem into the required form 
(see \cref{app:mean-field}).

For later convenience, we characterise the driving field by
its cumulant generating functional:
\begin{multline}
  \label{eq:src-cumulant-generating-functional}
  W[iJ]
  = \ln \vev{e^{i \int  J_a \eta_a \dd \bm{x} \dd t}}_{\eta}\, \\
  = \sum_{n=1}^\infty \frac{i^n}{n!}\,
  \int\!\dd\bm{x}_1 \int\!\dd t_1 \cdots \int\!\dd\bm{x}_n  \int\!\dd t_n\; \\
  \cumulant_{a_1\cdots a_n}(\bm{x}_1, t_1; \ldots; \bm{x}_n, t_n) \\
  J_{a_1}(\bm{x}_1, t_1)\cdots J_{a_n}(\bm{x}_n, t_n)\,,
\end{multline}
where $\vev{X}_{\eta} = \int X[\eta] \, P[\eta] \, \mathcal{D}\eta$
denotes the expectation value over driving-field realisations, $P[\eta]$
is the probability distribution functional of $\eta$, and $\mathcal{D}$
denotes the functional integral (integration over the field variable at
each point in $\bm{x}$ and $t$). Here $J_a$ is an auxiliary current
conjugate to $\eta_a$, and $\cumulant_{a_1\cdots a_n}$ are the
$n$th-order cumulants of the driving field. The driving field is
Gaussian if $W$ truncates at second order, i.e.\
$\cumulant_{a_1\cdots a_n} = 0$ for all $n > 2$. We focus on the
case of zero-mean driving ($\cumulant_a = 0$); non-zero mean can be
accommodated by performing a mean-field subtraction.

Since each realisation of $\eta$ determines a unique field
configuration $\varphi$ via \cref{eq:langevin}, the probability
distribution of $\varphi$ can be written as a path integral over
driving-field realisations:
\begin{equation}
  \begin{split}
    P[\varphi] &= \int \mathcal{D}\eta\; P[\eta] \;
    \delta^N \! \left[\partial_t \varphi - A \varphi - \mathcal{F}[\varphi] - \eta\right] \\
  \end{split}
  \label{eq:delta-functional}
\end{equation}
where the $N$-dimensional delta functional enforces \cref{eq:langevin}
for each field component at every point (indices are suppressed for
brevity). The functional probability $P[\varphi]$ is not normalised by
construction; expectation values are normalised explicitly by the
partition function in \cref{eq: average O with S}. A closer look at the normalisation shows what
\cref{eq:delta-functional} omits: the Jacobian
$J[\varphi] = \operatorname{Det}(\delta f/\delta\varphi)$\footnote{A
functional determinant is defined through a discretisation of time and
space, under which $\delta f/\delta\varphi$ is a finite matrix; writing it
implicitly performs that discretisation, whose time-slicing convention
fixes $\Theta(0)$ (\cref{sec:ito}).} of the map
$\varphi \mapsto f[\varphi] = \partial_t\varphi - A\varphi -
\mathcal{F}[\varphi]$, which relates the delta functional of the equation
to that of the solution,
$\delta[\varphi - \varphi_\eta] = J[\varphi]\,\delta[f[\varphi] - \eta]$.
In the \ito{} (pre-point) discretisation adopted throughout, $J$ is
field-independent (and equal to one in the time-sliced integral), so it
affects only the normalisation; in the Stratonovich (mid-point)
discretisation $J$ depends on $\varphi$ and must be kept in the action
\cite{Hertz2017,ChowBuice2015}. \Cref{app:jacobian} gives the derivation.

Representing the delta functional via its Fourier transform introduces
conjugate response fields $\psi_a(\bm{x}, t)$ (one for each physical
field component), following Martin, Siggia, and Rose
\cite{Martin1973,Janssen1976,DeDominicis1976}:
\begin{equation}
  \begin{split}
    P[\varphi] &= \int \mathcal{D}\psi \int \mathcal{D}\eta\; P[\eta] \\
    &\quad \exp\Bigl[- i \int \psi_a(\bm{x}, t) \bigl(\partial_t \varphi_a(\bm{x}, t) - A_{a b} \varphi_b(\bm{x}, t) \\
    &\qquad - \mathcal{F}_a[\varphi] - \eta_a(\bm{x}, t)\bigr) \dd \bm{x} \dd t\Bigr] \\
    & = \int \mathcal{D}\psi \;
    \exp\Bigl[- i \int \psi_a(\bm{x}, t) \bigl(\partial_t \varphi_a(\bm{x}, t) - A_{a b} \varphi_b(\bm{x}, t) \\
    &\qquad - \mathcal{F}_a[\varphi] \bigr) \dd \bm{x} \dd t\Bigr]\,
    e^{W[i \psi]}
  \end{split}
  \label{eq: physical field prob}
\end{equation}
where the second equality follows from integrating out $\eta$ and
recognising the result as the cumulant generating functional $W[i\psi]$
(\cref{eq:src-cumulant-generating-functional}).

The probability distribution is expressed more compactly through an
effective action that encodes both the deterministic dynamics and the
driving-field statistics:
\begin{equation}
    P[\varphi]
    \equiv \int \mathcal{D}\psi \;
    e^{- S[\varphi,\psi]} ,
  \label{eq: physical field prob 2}
\end{equation}
where we define the MSR action
\begin{multline}
  S[\varphi,\psi] = i \int \psi_a(\bm{x}, t) \bigl(\partial_t \varphi_a(\bm{x}, t) - A_{a b}(\bm{x}, t) \varphi_b(\bm{x}, t) \\
  - \mathcal{F}_a[\varphi] \bigr) \dd \bm{x} \dd t - W[i \psi]\,.
\end{multline}
This is the MSR action for a multi-component field theory with
arbitrary driving-field statistics, and serves as the starting point
for the perturbative expansion developed in the next subsection.

\subsection{Observables and perturbative expansion}

The expectation value of an observable $O[\varphi]$ is computed as a
path integral weighted by the MSR action:
\begin{equation}
  \label{eq: average O with S}
  \vev{O[\varphi]}_S = \frac{1}{Z_S}
  {\int \mathcal{D}\varphi\,\mathcal{D}\psi\; O[\varphi] e^{-S[\varphi,\psi]}}
  \,,
\end{equation}
where $Z_S=\int \mathcal{D}\varphi\,\mathcal{D}\psi\; e^{-S[\varphi,\psi]}
= \int\mathcal{D}\varphi\,P[\varphi]$ is the partition function. The
division by $Z_S$ is the normalisation referred to after
\cref{eq:delta-functional}: a field-independent constant dropped there
cancels between numerator and denominator. With the measure normalised on
the linear theory and the \ito{} prescription, $Z_S = 1$ exactly for any
$\mathcal{F}$ (\cref{app:ZS-unity}), so the normalisation is trivial and
the perturbative series below needs no correction at any truncation order.

Direct evaluation of the path integral is generally intractable. To
proceed perturbatively, we decompose the MSR action as
\begin{equation}
  \label{eq:action-decomposition}
  S[\varphi,\psi] = S_0[\varphi,\psi] + \Sint[\varphi,\psi]\,,
\end{equation}
where the free (Gaussian) part $S_0$ captures the linear dynamics and
Gaussian driving-field statistics,
\begin{multline}
  S_0 = i \int \psi_a(\bm{x}, t) \bigl(\partial_t \varphi_a(\bm{x}, t) - A_{a b}(\bm{x}, t) \varphi_b(\bm{x}, t) \bigr) \dd \bm{x} \dd t \\
  - W_{\mathrm{G}}[i \psi]\,,
\end{multline}
with $W_{\mathrm{G}}$ containing the first- and second-order cumulants
(mean and variance) of the driving field, while the interaction part
$\Sint$ collects the deterministic nonlinearities and non-Gaussian
driving-field cumulants:
\begin{equation}
  \Sint = - i \int \psi_a(\bm{x}, t) \mathcal{F}_a[\varphi] \dd \bm{x} \dd t - W_{\mathrm{NG}}[i \psi]\,,
\end{equation}
where $W_{\mathrm{NG}}$ includes the third and higher order cumulants convolved with the response fields.

The perturbative expansion treats $\Sint$ as a perturbation to the free
theory:
\begin{equation}
  \label{eq:perturbative}
  \vev{O}_S =
  \sum_{n=0}^{N} \frac{(-1)^n}{n!}\,\vevfree{O\,\Sint^n}
  \, ,
\end{equation}
where $\vevfree{\cdots}$ denotes the expectation value with respect to
$S_0$. 

The expectation values of observables of interest, typically correlation 
functions of the physical fields, can be computed from the two-point functions 
(propagators) of $S_0$ using Wick's theorem, given that $S_0$ is Gaussian 
and $\Sint$ is polynomial in $\varphi$ and $\psi$.
The propagators are defined in
\cref{sec:propagators} and Wick's theorem is reviewed in
\cref{sec:wick-theorem}.

\subsection{Free-theory propagators}
\label{sec:propagators}

The observables of primary interest are correlation functions (moments)
of the physical fields, in which case $O$ is a product of $\varphi$
fields at specified spacetime points. To compute these systematically,
we introduce the moment generating functional for the free theory by
coupling auxiliary currents $J_a(\bm{x}, t)$ and $K_a(\bm{x}, t)$ to
the physical and response fields, respectively:
\begin{multline}
  \mathcal{Z}_0[J,K] = \int \mathcal{D}\varphi\,\mathcal{D}\psi\;
  e^{-S_0[\varphi,\psi]} \\
  e^{\int J_a(\bm{x}, t) \varphi_a(\bm{x}, t)
  \dd \bm{x} \dd t} e^{\int K_a(\bm{x}, t) \psi_a(\bm{x}, t) \dd \bm{x} \dd t}\,.
\end{multline}
Free-theory moments are obtained by differentiation with respect to
$J$ and $K$. For example,
\begin{equation}
    \left\langle
    \varphi_{a}(\bm{x}, t)\psi_{b}(\bm{x}', t')
    \right\rangle_{S_0}
    =
    \frac{\delta^2 \mathcal{Z}[J, K]}{\delta J_{a} (\bm{x}, t)
    \delta K_{b} (\bm{x}', t')
    }
    \Big\vert_{J, K = 0}
    \label{eq: moment example}
\end{equation}
where differentiation with respect to $J$ brings down the
corresponding $\varphi$, and differentiation with respect to $K$
brings down the corresponding $\psi$.

The free action $S_0$ is quadratic in the fields and can be written in
the bilinear form
\begin{multline}
    \label{eq: rewriting S_0 continuous}
    S_0[\varphi, \psi] =
    i \int \dd{\bm{x}} \dd{\bm{x}'}\dd{t} \dd{t'} \;
    \psi_{a}(\bm{x}', t') \\
    \DeterOp_{a b}(\bm{x}', t';  \bm{x}, t)
    \varphi_{b}(\bm{x}, t) \\
    +
    \frac{1}{2} \int
        \dd{\bm{x}} \dd{\bm{x}'}\dd{t} \dd{t'} \, \psi_{a}(\bm{x}, t) \\
        \cumulant_{a b}(\bm{x}, t; \bm{x}', t') \, \psi_{b}(\bm{x}', t') \,,
\end{multline}
where $\DeterOp$ is the linear-dynamics operator:
\begin{equation}
\DeterOp_{a b}(\bm{x}', t';  \bm{x}, t) \equiv
\delta(\bm{x}' - \bm{x}) \,
\delta(t' - t) \,
\left[\delta_{a b} \partial_t - A_{a b}(\bm{x}, t)\right].
\end{equation}
Upon discretising the field coordinates, \cref{eq: rewriting S_0 continuous}
takes the standard matrix form
\begin{equation}
    S_0 = \frac{1}{2}
    \begin{pmatrix}
        \varphi^T & \, \psi^T
    \end{pmatrix}
    \begin{pmatrix}
        0 & \, i \DeterOp^T \\[3pt]
        i \DeterOp & \,\cumulant
    \end{pmatrix}
    \begin{pmatrix}
        \varphi \\[3pt]
        \psi
    \end{pmatrix}
    \label{eq: action ref quadratic}
\end{equation}
where each matrix product implicitly represents integration over
$\bm{x}$ and $t$ together with summation over component indices.

The propagator matrix is defined as the inverse of the quadratic form
in \cref{eq: action ref quadratic}:
\begin{equation}
    G_0
    =
    \begin{pmatrix}
        0 & \, i\DeterOp^T \\[3pt]
        i\DeterOp & \;\cumulant
    \end{pmatrix}^{-1}
    =
    \begin{pmatrix}
        \CorrOp & -i \, \RespOp \\[3pt]
        -i \, \RespOp^T &  0
    \end{pmatrix}
\end{equation}
where $\RespOp\equiv \DeterOp^{-1}$ is the \textit{response} propagator
and $\CorrOp \equiv \RespOp \cumulant\RespOp^T$ is the
\textit{correlation} propagator (the covariance of the linearly
propagated driving field, as \cref{eq: Corr Op} below makes explicit). Indices and arguments are suppressed
for compactness.

Evaluating $\mathcal{Z}_0$ as a Gaussian integral yields
\begin{equation}
\begin{split}
    \ln{\mathcal{Z}_0[J, K]}
    & =
    \frac{1}{2}
    \begin{pmatrix}
        J^T & \, K^T
    \end{pmatrix}
    \begin{pmatrix}
        \CorrOp & -i \, \RespOp \\[3pt]
        -i \, \RespOp^T &  0
    \end{pmatrix}
    \begin{pmatrix}
        J \\[3pt]
        K
    \end{pmatrix} \\
    & =
    \frac{1}{2}J^T \CorrOp J - i\, J^T \RespOp K ,
\end{split}
\end{equation}
where the normalisation constant has been dropped.
Using \cref{eq: moment example} and converting back to continuous notation,
the two-point functions are obtained as follows:
\begin{align}
    \left\langle
    \varphi_{a}(\bm{x}, t)\psi_{b}(\bm{x}', t')
    \right\rangle_{S_0}
    & \equiv
    - i \,
    \RespOp_{a b}(\bm{x}, t; \bm{x}', t') \,,\\
    \left\langle
    \varphi_{a}(\bm{x}, t)\varphi_{b}(\bm{x}', t')
    \right\rangle_{S_0}
    & \equiv
    \CorrOp_{a b}(\bm{x}, t; \bm{x}', t') \,,
    \label{eq: moment example 2}
\end{align}
which makes the naming transparent: $\CorrOp$ is the autocorrelation of
the physical fields, while $\RespOp$ characterises the linear response
of $\varphi$ to perturbations coupled to $\psi$.
The response-field autocorrelation vanishes identically:
\begin{equation}
  \vevfree{\psi_a(\bm{x}, t)\,\psi_b(\bm{x}', t')}= 0
    \,.\label{eq:psi-psi-zero}
\end{equation}
The vanishing of the $\psi$--$\psi$ correlator,
\cref{eq:psi-psi-zero}, is a structural property of the MSR formalism
that distinguishes it from standard quantum field theory.

\paragraph{Relation to the Schwinger--Keldysh formalism}
The structure of \cref{eq: moment example 2,eq:psi-psi-zero} is the
classical limit of the closed-time-path (Schwinger--Keldysh) formalism
of quantum dynamics \cite{Kamenev2023,Altland2010}. There, every degree
of freedom is doubled into fields $\phi_{+}$ and $\phi_{-}$ on the forward
and backward branches of the contour, and the Keldysh rotation defines a
``classical'' component $\phi_{\mathrm{cl}} \propto \phi_{+} + \phi_{-}$
and a ``quantum'' component $\phi_{\mathrm{q}} \propto \phi_{+} - \phi_{-}$.
In the classical limit the action is expanded to quadratic order in
$\phi_{\mathrm{q}}$, and the MSR action of \cref{sec:msr-action} with
Gaussian driving follows with $\varphi \leftrightarrow \phi_{\mathrm{cl}}$
and $\psi \leftrightarrow \phi_{\mathrm{q}}$; the higher cumulant vertices
of \cref{sec:example2} correspond to higher powers of $\phi_{\mathrm{q}}$.
The propagator matrix $G_0$ is the
Keldysh-rotated Green's function: $\CorrOp$ is the Keldysh (symmetric)
component, $\RespOp$ and $\RespOp^{T}$ are the retarded and advanced
components, and the vanishing of the quantum--quantum component,
$G^{qq} = 0$, is \cref{eq:psi-psi-zero}. Likewise, the
normalisation $Z_S = 1$ (\cref{app:ZS-unity}) is the MSR counterpart of
the Keldysh normalisation $Z = 1$, which follows from the closed time
contour and is the reason no vacuum diagrams need to be subtracted in
either formalism.


\subsection{Explicit expressions for propagator operators}

The response propagator, as the inverse of the linear operator $\DeterOp$, also known as the Green's function, satisfies
\begin{multline}
\int \dd{\bm{x}}\dd{t} \,
\DeterOp_{a b}(\bm{x}', t'; \bm{x}, t) \,
\RespOp_{b c}(\bm{x}, t; \bm{x}'', t'') \\
=
\delta_{a c} \, \delta(\bm{x}' - \bm{x}'') \, \delta(t' - t''),
\end{multline}
with the explicit solution\footnote{
  The temporal part of the response operator is the solution of the differential equation
  $\frac{d}{dt} U(t,t') = A(t)\, U(t,t'), \quad U(t',t') = I$.
}
\begin{multline}
    \RespOp_{a b}(\bm{x}, t; \bm{x}', t')
    =
    \delta(\bm{x} - \bm{x}')  \,
    \Theta(t - t') \\
    \left\{ \mathcal{T} \! \exp \left( \int_{t'}^t \mathbf{A}(\bm{x}, \tau) \dd{\tau} \right) \right\}_{a b},
    \label{eq: ResPon Op}
\end{multline}
where $\Theta$ is the Heaviside step function\footnote{We adopt the
\ito{} convention, in which $\Theta(0) = 0$ so that the response
propagator is strictly retarded; see \cref{sec:ito}.}
\begin{equation}
\Theta(\Delta t) = \begin{cases}
0 & \text{if } \Delta t \leq 0, \\
1 & \text{if } \Delta t > 0,
\end{cases}
\end{equation}
and `$\mathcal{T}\! \exp$' represents the time-ordered exponential, defined as a Dyson series expansion:
\begin{multline}
\mathcal{T}\!\exp\!\left(\int_{t'}^{t} \mathbf{A}(\tau)\,d\tau\right)
=
\mathbf{1}
+ \int_{t'}^{t} d\tau_1\, \mathbf{A}(\tau_1) \\
+ \int_{t'}^{t} d\tau_1 \int_{t'}^{\tau_1} d\tau_2\, \mathbf{A}(\tau_1)\mathbf{A}(\tau_2) \\
+ \int_{t'}^{t} d\tau_1 \int_{t'}^{\tau_1} d\tau_2 \int_{t'}^{\tau_2} d\tau_3\,
\mathbf{A}(\tau_1)\mathbf{A}(\tau_2)\mathbf{A}(\tau_3)
+ \cdots  ,
\end{multline}
where operators at later $\tau$ appear to the left. 

The correlation propagator is given by contracting (or convolving) two response functions with a two-point cumulant of driving fields:
\begin{multline}
    \CorrOp_{a b}(\bm{x}_1, t_1; \bm{x}_2, t_2)
    = \int \dd{\bm{x}'}\dd{\bm{x}''}\dd{t'}\dd{t''} \\
     \RespOp_{a d}(\bm{x}_1, t_1; \bm{x}', t') \,
    \cumulant_{d c}(\bm{x}', t'; \bm{x}'', t'') \,
    \RespOp_{b c}(\bm{x}_2, t_2; \bm{x}'', t'') .
    \label{eq: Corr Op}
\end{multline}
Substituting \cref{eq: ResPon Op} into the above equation yields
\begin{multline}
    \CorrOp_{a b}(\bm{x}_1, t_1; \bm{x}_2, t_2)
    = \int \dd{t'}\dd{t''}
    \cumulant_{d c}(\bm{x}_1, t'; \bm{x}_2, t'') \, \\
    \left\{ \mathcal{T}\! \exp \left( \int_{t'}^{t_1} \mathbf{A}(\bm{x}_1, \tau) \dd{\tau} \right) \right\}_{a d} \, \\
    \left\{ \mathcal{T}\! \exp \left( \int_{t''}^{t_2} \mathbf{A}(\bm{x}_2, \tau) \dd{\tau} \right) \right\}_{bc}\,.
    \label{eq: Corr Op 2}
\end{multline}
To conclude, we have obtained explicit expressions for the two propagators 
in the free-field theory: the response operator $\RespOp$, which encodes the linear 
deterministic  evolution, and the correlation operator $\CorrOp$, which describes 
the propagation of Gaussian statistics of the driving fields.
The correlation propagator $\CorrOp$
is symmetric:
$C_{ab}(\bm{x}, t;\bm{x}', t') = C_{ba}(\bm{x}', t'; \bm{x}, t)$.
The response propagator $\RespOp$ is retarded (causal):
$R_{ab}(\bm{x}, t; \bm{x}', t') = 0$ whenever $t < t'$, reflecting
the physical requirement that the system cannot respond to future
perturbations. 
The propagators capture only the
dynamics generated by the driving field from $t_{\min}$ onward and
carry no information about the initial state. The role of initial
conditions and how they are handled, including their elimination via
mean-field subtraction, is discussed in \cref{app:initial-conditions}.

\paragraph{Special Cases}

Equations~(\ref{eq: ResPon Op}) and (\ref{eq: Corr Op 2}) are the most general forms of the propagators.
However, most applications often admit simplifications. 
In a special (and most common) case where the linear operator is diagonal, 
$\mathbf{A}_{ab}(t)=\delta_{ab} \lambda_a(t)$, then
\begin{equation}
  \left\{ \mathcal{T} \! \exp \left( \int_{t'}^t \mathbf{A}(\bm{x}, \tau) \dd{\tau} \right) \right\}_{a b}
  =
  \delta_{ab}\,\exp\!\left(\int_{t'}^{t}\lambda_a(\tau)\,d\tau\right),
\end{equation}
so each component evolves independently. The response operator reduces to
ordinary exponentials.

On the other hand, in the special case of statistically homogeneous
and isotropic driving fields, the cumulant functions admit an expansion
in a suitable basis. 
For example, on the sphere $S^2$ with angular coordinates $\hat{n}$,
\begin{equation}
\cumulant_{ab}(\hat n_1,t';\hat n_2,t'')
=
\sum_{\ell}
\frac{2\ell+1}{4\pi}
C_\ell^{ab}(t',t'')
P_\ell(\hat n_1\!\cdot\!\hat n_2)  ,
\end{equation}
which in turn yields a significantly simplified form for the correlation propagator.

\subsection{Wick's theorem in the MSR formalism}
\label{sec:wick-theorem}

Since the free theory is Gaussian, expectation values of products of
fields can be evaluated using Wick's theorem: the $2n$-point moment of
centred Gaussian variables decomposes into a sum over all complete
pairwise contractions (products of two-point functions).

Concretely, for jointly Gaussian random variables $X_1, X_2, \ldots, X_{2n}$ with zero mean,
\begin{equation}
  \label{eq:wick}
  \langle X_1 X_2 \cdots X_{2n} \rangle
  =
  \sum_{p \in P_n^2} \prod_{\{i,j\} \in p} \langle X_i X_j \rangle\,,
\end{equation}
where $P_n^2$ is the set of all partitions of $\{1, \ldots, 2n\}$ into
$n$ unordered pairs. The number of such partitions is
\begin{equation}
(2n-1)!! = \frac{(2n)!}{2^n n!}\,,
\end{equation}
which grows super-exponentially with $n$. As an example, for a
four-point function ($2n=4$) there are $(4-1)!! = 3$ terms:
\begin{multline}
\langle X_1 X_2 X_3 X_4 \rangle = \langle X_1 X_2 \rangle \langle X_3 X_4 \rangle \\ + \langle X_1 X_3 \rangle \langle X_2 X_4 \rangle  + \langle X_1 X_4 \rangle \langle X_2 X_3 \rangle.
\end{multline}
In the present context, each $X$ is either a $\varphi$ or $\psi$ field
operator.

In the MSR formalism, many pairings vanish by virtue of
\cref{eq:psi-psi-zero}: any pairing that includes a $\psi$--$\psi$
contraction contributes zero. Only the propagators $\CorrOp$ (from
$\varphi$--$\varphi$ pairings) and $\RespOp$ (from $\varphi$--$\psi$
pairings) survive. In particular,
\begin{equation}
    \left\langle
    \prod_{i=1}^{n_1} \prod_{j=1}^{n_2}\varphi_{a_i} \psi_{b_j}
    \right\rangle_{S_0}
    =
    0,
    \qquad
    \text{if } n_2 > n_1.
\end{equation}
This MSR constraint significantly reduces the number of valid pairings,
but the count remains combinatorially large at higher perturbative
orders. Rather than relying on brute-force enumeration, \sftwick{}
implements an efficient algorithm that identifies topologically distinct
pairings (characterised by Feynman diagrams) while enforcing all the 
vanishing conditions; the algorithm is described in \cref{sec:contraction}.

\subsection{The \ito{} prescription}
\label{sec:ito}

The discretisation of the stochastic integral in \cref{eq:langevin}
introduces an ambiguity in the equal-time limit of the response
propagator. In the \ito{} convention \cite{Gardiner2009}, the driving
field is evaluated at the beginning of each time step, which translates
to
\begin{equation}
  \label{eq:ito}
  \Theta(0) = 0 \quad\Longrightarrow\quad
  R_{ab}(\bm{x}, t ; \bm{x}', t) = 0\,.
\end{equation}
That is, the equal-point response propagator vanishes identically,
eliminating any Wick contraction that pairs a $\varphi$ and a $\psi$ at
the same spacetime point and thereby further pruning the enumeration.


\subsection{Causal structure and retarded-propagator loops}
\label{sec:causal}

Since $\RespOp$ is retarded, each response propagator defines a directed
edge from the $\psi$-endpoint to the $\varphi$-endpoint in the causal
ordering: $R_{ab}(\bm{x}, t ; \bm{x}', t')$ requires $t \geq t'$. A
directed cycle in the response propagator graph is therefore causally
forbidden. For example, a two-cycle
$R(\bm{x}, t; \bm{x}', t')\,R(\bm{x}', t'; \bm{x}'', t'')$ requires
both $t' > t''$ and $t'' > t'$, which is never satisfied. More
generally, any $n$-cycle
\begin{equation}
        \RespOp(\bm{x}, t_1;\bm{x},  t_2)\RespOp(\bm{x}, t_2;\bm{x},  t_3) 
        \cdots \RespOp(\bm{x}, t_n; \bm{x}, t_1)=0
\end{equation}
requires $t_1 \geq t_2 \geq \cdots \geq t_n \geq t_1$, forcing all
times equal, which vanishes under the \ito{} rule.

The elimination of causal $\RespOp$-loops is a non-trivial constraint that
removes many otherwise valid Wick pairings at higher orders. In
\sftwick{}, this is implemented via directed-graph cycle detection
(depth-first search) on the $\RespOp$-edge subgraph of each candidate
pairing.

\subsection{Feynman diagram representation}
\label{sec:feynman-diagrams-theory}

Each term in the Wick expansion corresponds to a Feynman diagram, a
multigraph whose nodes represent external points (observable field
insertions) and internal vertices (from $\Sint$), and whose edges
represent propagators.

\paragraph{Edges}
The propagator types are visually distinguished as follows:
\begin{itemize}
  \item $\CorrOp$ propagators are drawn as undirected edges (solid lines),
  \item $\RespOp$ propagators are drawn as directed edges (dashed lines with
    arrows), pointing from the $\psi$-endpoint to the
    $\varphi$-endpoint. Since
    $\vevfree{\varphi_a\,\psi_b} = -\ii \, \RespOp_{ab}$, each $\RespOp$ edge
    carries a phase factor $(-\ii)$; a diagram containing $n_R$
    response propagators therefore acquires an overall phase
    $(-\ii)^{n_R}$.
\end{itemize}

\paragraph{Vertices}
Each monomial term in $\Sint$ defines an interaction vertex, and the
perturbative expansion sums over all ways of connecting $n$ vertices
to the observable $O$ via free-theory propagators. When the interaction
action consists of multiple vertex types,
$\Sint = \sum_\alpha V_\alpha$, the $n$th power expands via the
multinomial theorem:
\begin{equation}
  \Sint^n = \sum_{\substack{n_1,\ldots,n_k \geq 0 \\ n_1+\cdots+n_k = n}}
  \frac{n!}{n_1!\cdots n_k!}\,V_1^{n_1}\cdots V_k^{n_k}\,.
\end{equation}
Combined with the $1/n!$ prefactor in \cref{eq:perturbative}, the net
coefficient for a given combination $(n_1,\ldots,n_k)$ becomes
$(-1)^n/(n_1!\cdots n_k!)$.

Each vertex $V_\alpha$ is defined by a set of fields and a coupling
tensor. For a local cubic vertex, for instance,
\begin{equation}
  V = - \ii \int\!\dd\bm{x}\dd t\; F^{\mathrm{(3)}}_{abc}
  \,\psi_a(\bm{x}, t)\,\varphi_b(\bm{x}, t)\,\varphi_c(\bm{x}, t)\,,
\end{equation}
where $F^{\mathrm{(3)}}_{abc}$ is the coupling tensor and the integral runs over
all spacetime points. A vertex may also be non-local, for example in terms arising 
from $W_{\mathrm{NG}}$, which involve fields evaluated at different spatial points.

Wick contractions decompose higher-order correlations into products of propagators, 
with each internal vertex corresponding to an integral over the intermediate field coordinates.
Different Wick pairings can lead to topologically identical Feynman diagrams: 
they share the same propagator connectivity, but differ in how component indices 
are routed through the diagram. Identifying and combining such equivalent contributions 
is a key task performed by the diagram-processing pipeline of \sftwick{}
(\cref{sec:simplification}).

\section{Worked example I: Gaussian driving field}
\label{sec:examples}


We demonstrate \sftwick{} by computing the two-point correlation function
of an example two-component, one-dimensional field. The system
includes up to quadratic nonlinearities (cubic vertices in the MSR action)
and is driven by Gaussian stochastic noise. This setup represents a common
configuration within our formalism: local interactions
with only the two-point noise cumulant non-vanishing.
This example features non-trivial multi-component coupling and an additional
spatial dimension, providing a sufficient test of the generalised formalism
and its core code implementation.


Less abstractly, we consider the following coupled Langevin equations
\begin{align}
  \partial_t\varphi_1(x,t) &= -\varphi_1(x,t)
    + \varphi_2^2(x,t) + \eta_1(x,t)\,,\\
  \partial_t\varphi_2(x,t) &= -\varphi_2(x,t)
    + \varphi_1(x,t)\varphi_2(x,t) + \eta_2(x,t)\,,
\end{align}
where the real, stochastic driving fields are Gaussian with zero mean
and factorised space--time covariance
\begin{align}
  \cumulant_{11}(x,t;x',t') &= \cumulant_{22}(x,t;x',t')
    = \lambda\,e^{-|t-t'|/\sigma_t}\,e^{-|x-x'|/\sigma_x}\,,
    \label{eq:kappa-example}\\
  \cumulant_{12}(x,t;x',t') &= \cumulant_{21}(x,t;x',t') = 0\,,
\end{align}
with $\cumulant_{ab}(x,t;x',t') \equiv
\langle\eta_a(x,t)\,\eta_b(x',t')\rangle$. The linear operator is
$A_{ab} = -\delta_{ab}$, giving the causal response and correlation
propagators
\begin{align}
  \RespOp_{ab}(x,t;x',t')
    &= \delta(x-x')\,\Theta(t-t')\,e^{-(t-t')}\,\delta_{ab}\,,
    \label{eq: ResPon Op example}\\
  \CorrOp_{ab}(x,t;x',t')
    &= e^{-|x-x'|/\sigma_x}\int_{0}^{t}\dd{\tau}\int_{0}^{t'}\dd{\tau'}\nonumber\\
    &\qquad \lambda\,e^{-|\tau-\tau'|/\sigma_t}\,
       e^{-(t-\tau)}\,e^{-(t'-\tau')}\,\delta_{ab}\,.
    \label{eq: Corr Op example}
\end{align}
The coupling tensor $F^{\mathrm{(3)}}_{abc}$ has two independent
non-zero entries: $F^{\mathrm{(3)}}_{122} = 1$ (the $\varphi_2^2$
source in the first equation) and
$F^{\mathrm{(3)}}_{212} = F^{\mathrm{(3)}}_{221} = \tfrac{1}{2}$
(the symmetrised $\varphi_1\varphi_2$ source in the second).
We adopt the parameter values $\lambda = 0.05$ (weak coupling, so that
successive orders are suppressed), $\sigma_t = 0.3$, $\sigma_x = 1.0$, and $\gamma = 1$.
The exponential (Ornstein--Uhlenbeck) kernels in
\cref{eq:kappa-example} are chosen for their favourable numerical
properties: the resulting covariance matrices remain well-conditioned
at any grid spacing, the temporal noise can be generated exactly via
an autoregressive recursion, and the double integral in
\cref{eq: Corr Op example} admits a closed-form expression in terms
of elementary functions.


\begin{figure*}[t]
  \centering
  \includegraphics[width=\textwidth]{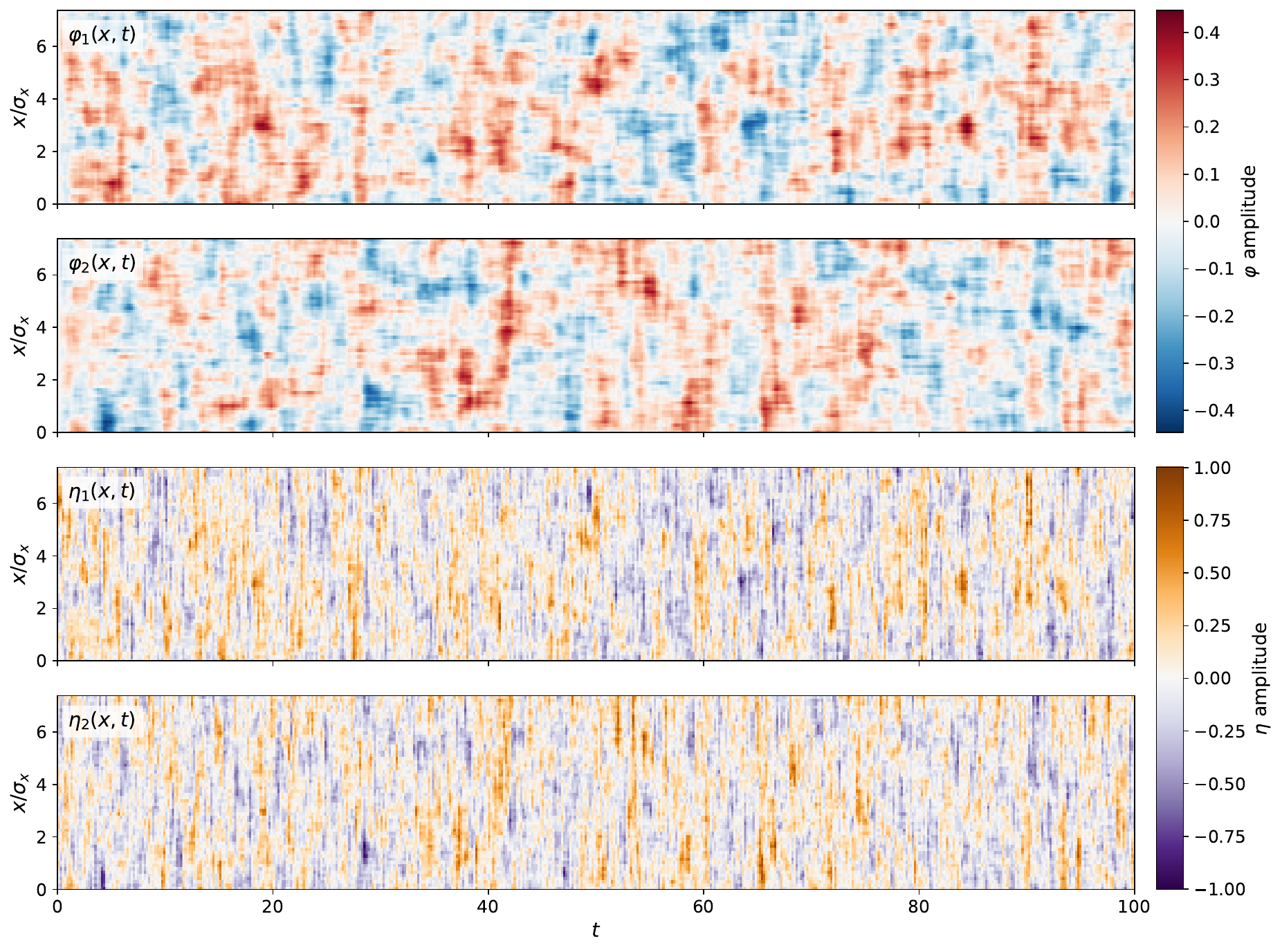}
  \caption{One realisation in space and time. \emph{Top}: the physical fields
    $\varphi_1$ and $\varphi_2$, starting from zero. \emph{Bottom}: the driving
    fields $\eta_1$ and $\eta_2$, with correlation length $\sigma_x$ and
    correlation time $\sigma_t = 0.3$. The two colour scales differ.}
  \label{fig:waterfall}
\end{figure*}

\subsection{Direct numerical simulation}
As a ground truth we solve the Langevin equations directly with a
Heun (predictor--corrector) integrator, which provides second-order
weak convergence at step size $\Delta t = 0.02$.
The temporally correlated noise is generated exactly via an
autoregressive recursion that exploits the Markov property of the
exponential kernel (see \cref{app:simulation}).
We accumulate $N = 100\,000$ independent realisations;
\cref{fig:waterfall} shows a single trajectory of both field
components.
The equal-time two-point correlator
$\xi_{ab}(r,t) = \langle\varphi_a(0,t)\,\varphi_b(r,t)\rangle$ is
estimated by averaging over the ensemble, with a
Monte~Carlo standard error $\lesssim\!0.6\%$
(\cref{eq:wick-noise}).


\subsection{Perturbative expansion and Feynman diagrams}
Using the path-integral formalism of \cref{sec:theory}, we compute
the equal-time two-point function up to fourth order:
\begin{equation}
  \vev{\varphi_a(x,t)\,\varphi_b(y,t)}_S
  \approx \sum_{n=0}^{4} \frac{(-1)^n}{n!}\,
  \vevfree{\varphi_a(x,t)\,\varphi_b(y,t)\,\Sint^n}\,,
\end{equation}
where the interaction action contains a single cubic vertex,
\begin{equation}
  \Sint = -\ii \int\!\dd x\,\dd t\;
  F^{(3)}_{abc}\,\psi_a(x,t)\,\varphi_b(x,t)\,\varphi_c(x,t)\,.
\end{equation}
Odd orders vanish because each cubic vertex contributes three fields,
so at order $n$ the total field count is $3n + 2$; for odd $n$ this
is odd, and Wick's theorem requires an even number of fields for a
complete pairing.

\begin{figure*}[t]
  \centering
  \includegraphics[width=0.45\textwidth]{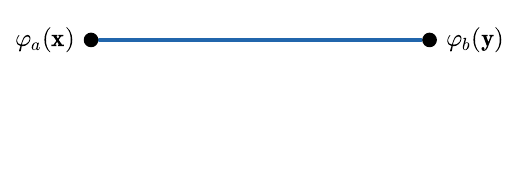}\\[6pt]
  \includegraphics[width=\textwidth]{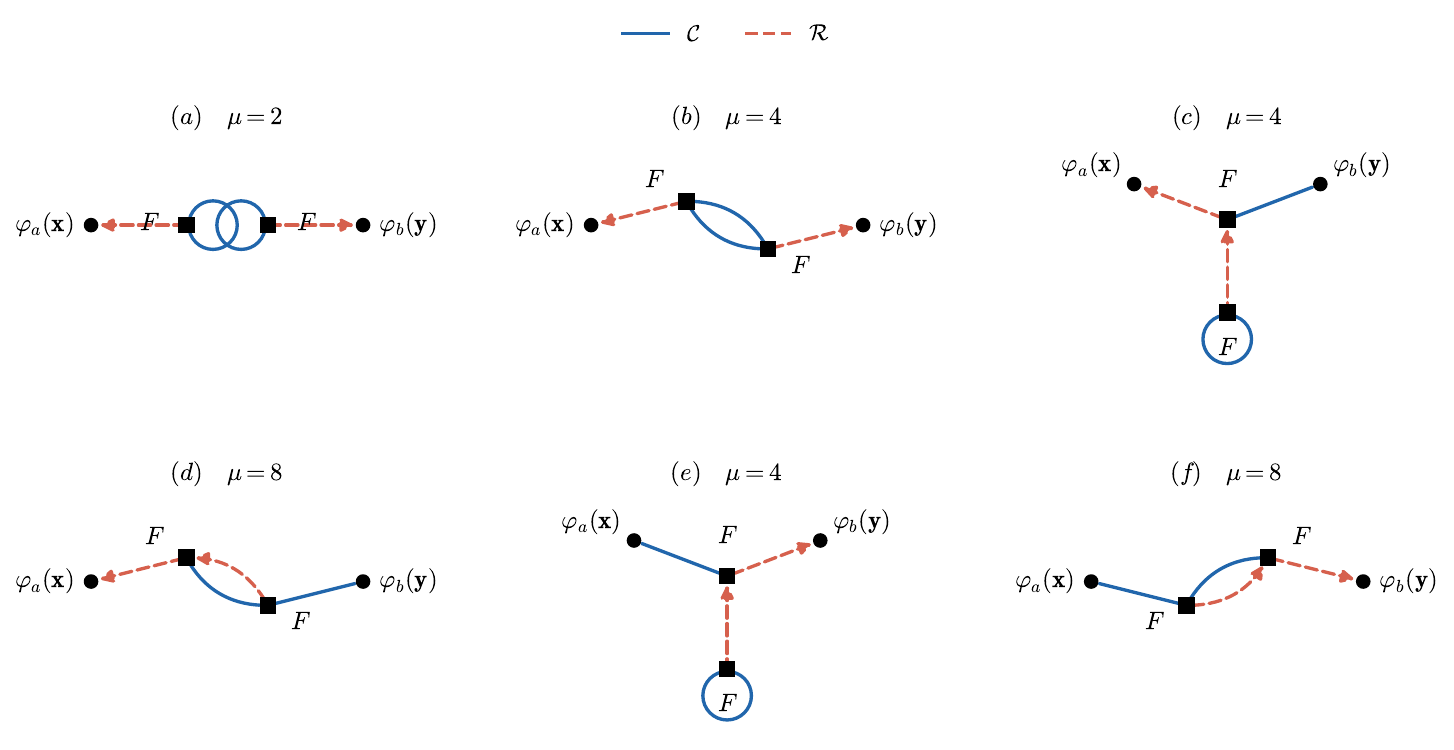}
  \caption{Diagrams for $\vev{\varphi_a(x,t)\,\varphi_b(y,t)}$ with the cubic vertex
    $(-\ii)\,F^{(3)}_{dab}\,\psi_d\varphi_a\varphi_b$. \emph{Top}: order zero.
    \emph{Bottom}: the six order-two topologies, with multiplicities $\mu$.
    Blue solid: $\CorrOp$; red dashed with arrow: $\RespOp$; circles: external
    points; squares: vertices.}
  \label{fig:example-diagrams}
\end{figure*}

At zeroth order,
$\vev{\varphi_a(x,t)\,\varphi_b(y,t)}^{(0)}
= \delta_{ab}\,\CorrOp(x,t;y,t)$
is a single $\CorrOp$-propagator (\cref{fig:example-diagrams}, top).
Because the propagators $\RespOp$ and $\CorrOp$ in this example are
diagonal and identical across components (isotropic), the component
indices reduce to a Kronecker delta prefactor and are suppressed
hereafter.

At second order, for generic coupling $F^{(3)}_{abc}$, \sftwick{}
identifies six topologically distinct diagrams
(\cref{fig:example-diagrams}, bottom). To declutter the expressions we
adopt the shorthand $\bm{x} = (x,t)$ for the external spacetime
points and $\bm{z}_i = (y_i,t_i)$ for the internal vertices, with
$\dd\bm{z}_i \equiv \dd y_i\,\dd t_i$.
Omitting coefficients and suppressing component indices,
diagrams~(a)--(f) correspond to the panels of
\cref{fig:example-diagrams} (bottom), read left to right, top to
bottom:
\begin{align}
  \text{(a)}\quad  &\int\!\dd\bm{z}_0\,\dd\bm{z}_1\;
    \RespOp(\bm{x};\bm{z}_0)\,\RespOp(\bm{y};\bm{z}_1)\,
    \CorrOp(\bm{z}_0;\bm{z}_0)\,\CorrOp(\bm{z}_1;\bm{z}_1)\,,
    \notag\\
  \text{(b)}\quad  &\int\!\dd\bm{z}_0\,\dd\bm{z}_1\;
    \RespOp(\bm{x};\bm{z}_0)\,\RespOp(\bm{y};\bm{z}_1)\,
    \CorrOp(\bm{z}_0;\bm{z}_1)\,\CorrOp(\bm{z}_0;\bm{z}_1)\,,
    \notag\\
  \text{(c)}\quad  &\int\!\dd\bm{z}_0\,\dd\bm{z}_1\;
    \RespOp(\bm{x};\bm{z}_0)\,\RespOp(\bm{z}_0;\bm{z}_1)\,
    \CorrOp(\bm{y};\bm{z}_0)\,\CorrOp(\bm{z}_1;\bm{z}_1)\,,
    \notag\\
  \text{(d)}\quad  &\int\!\dd\bm{z}_0\,\dd\bm{z}_1\;
    \RespOp(\bm{x};\bm{z}_0)\,\RespOp(\bm{z}_0;\bm{z}_1)\,
    \CorrOp(\bm{y};\bm{z}_1)\,\CorrOp(\bm{z}_0;\bm{z}_1)\,,
    \notag\\
  \text{(e)}\quad  &\int\!\dd\bm{z}_0\,\dd\bm{z}_1\;
    \RespOp(\bm{y};\bm{z}_0)\,\RespOp(\bm{z}_0;\bm{z}_1)\,
    \CorrOp(\bm{x};\bm{z}_0)\,\CorrOp(\bm{z}_1;\bm{z}_1)\,,
    \notag\\
  \text{(f)}\quad  &\int\!\dd\bm{z}_0\,\dd\bm{z}_1\;
    \RespOp(\bm{y};\bm{z}_0)\,\RespOp(\bm{z}_0;\bm{z}_1)\,
    \CorrOp(\bm{x};\bm{z}_1)\,\CorrOp(\bm{z}_0;\bm{z}_1)\,.
\end{align}
Every diagram contains exactly two $\RespOp$-propagators and two
$\CorrOp$-propagators: the two $\psi$ fields introduced by
$\Sint^2$ must each pair with a $\varphi$ (since
$\vev{\psi\,\psi}_{S_0} = 0$), producing two $\RespOp$-edges, and the
remaining four $\varphi$'s pair into two $\CorrOp$-edges.
In practice, the spatial Dirac delta $\delta(x - x')$ intrinsic to
each $\RespOp$-propagator (\cref{eq: ResPon Op example}) collapses 
the spatial integrals, thereby significantly reducing the effective 
dimensionality of each diagram integral.

At fourth order the number of topologically distinct diagrams grows
to~64 for generic $F^{(3)}_{abc}$, so we omit their explicit
expressions. Once the specific coupling values of this example are
substituted, many diagrams acquire vanishing coupling coefficients,
reducing the effective count. The surviving integrals are evaluated
with tensor-product Gauss--Legendre quadrature on the causally ordered
time simplex, yielding deterministic, seed-independent results.

The sweep behind the figures of the first version of this paper used a
quasi-Monte Carlo rule at $2^{15}$ Sobol points, although the text then
described Gauss--Legendre quadrature. The order-4 integrand
is concentrated within $\sim\!1/\gamma$ and $\sim\!\sigma_t$ of the upper
corner of a simplex whose side grows with $t$, so the fraction of samples
that reach it falls with $t$ and that rule is biased there rather than
merely noisy: at $t = 15$ the seed used sat $69\%$ above the
Gauss--Legendre value, and six scramblings scatter by $39$--$42\%$. The
node count of the Gauss--Legendre rule is set by the order-2 channel,
which needs the finer grid at large $t$; its residual error enters the
error budget below. Relative to the figures of the first version, the
plotted correlator moves by at most $1.7\%$ and the conclusions of this
section are unchanged.

\subsection{Results}

Since the noise is diagonal ($\cumulant_{12} = 0$) and the
propagators are isotropic, the cross-correlator $\xi_{12}$ vanishes
identically at all orders. We therefore focus on the two non-trivial
diagonal correlators,
$\xi_{11}(r,t) = \vev{\varphi_1(0,t)\,\varphi_1(r,t)}$ and
$\xi_{22}(r,t) = \vev{\varphi_2(0,t)\,\varphi_2(r,t)}$.

\begin{figure*}[t]
  \centering
  \includegraphics[width=\textwidth]{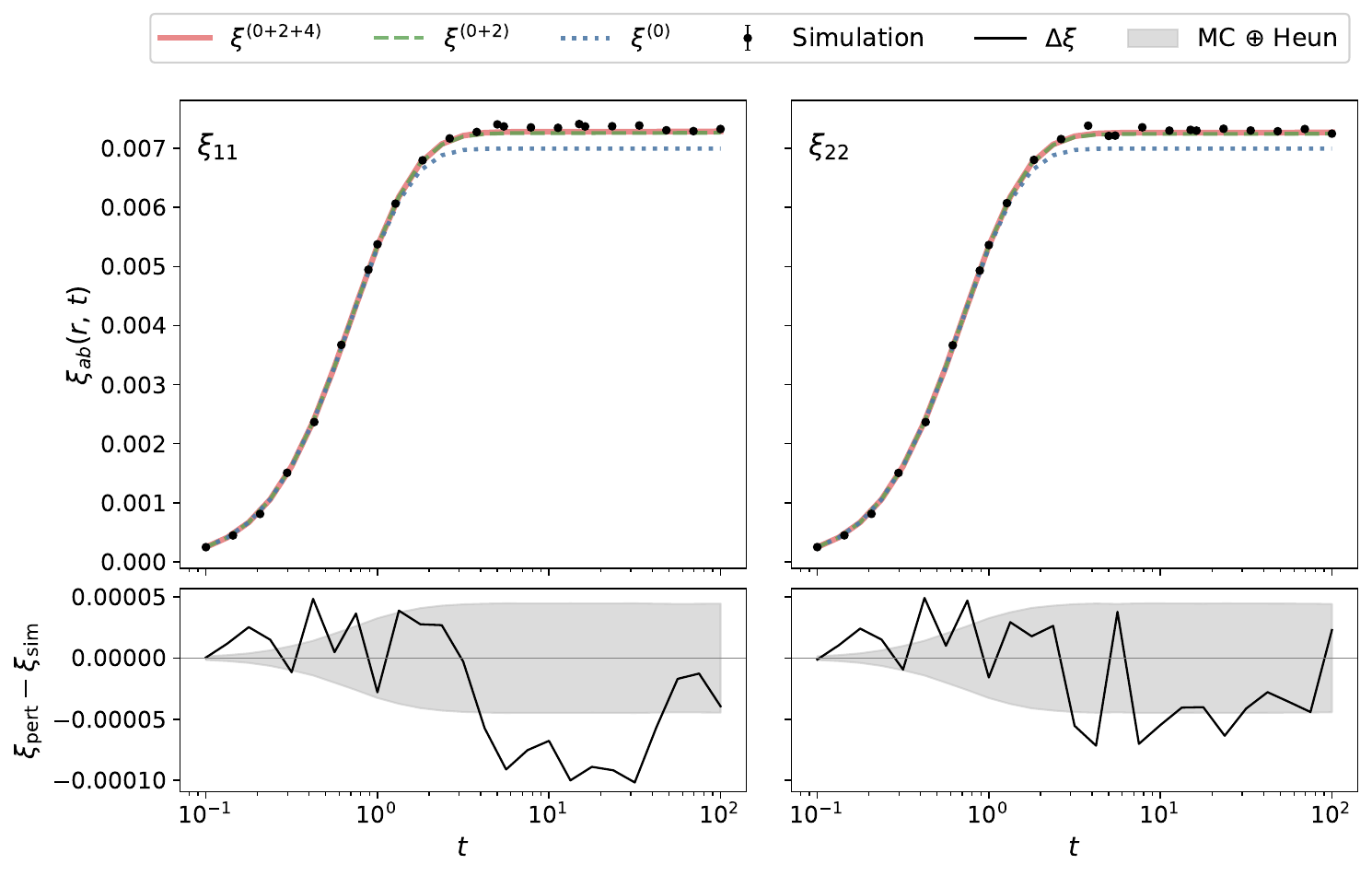}
  \caption{Two-point correlators at $r = 0.5\,\sigma_x$ versus time.
    \emph{Top}: perturbative predictions (dotted: $\xi^{(0)}$;
    dashed: $\xi^{(0+2)}$; solid: $\xi^{(0+2+4)}$) and simulation
    (markers with error bars). \emph{Bottom}: residual
    $\xi_{\mathrm{pert}} - \xi_{\mathrm{sim}}$ with combined
    uncertainty band. Left: $\xi_{11}$; right: $\xi_{22}$.}
  \label{fig:xi-vs-time}
\end{figure*}

\Cref{fig:xi-vs-time} shows the time evolution of both correlators
at the representative separation $r = 0.5\,\sigma_x$. Three
cumulative perturbative orders are plotted: the free theory
$\xi^{(0)}$ (dotted), the second-order correction $\xi^{(0+2)}$
(dashed), and the full fourth-order result $\xi^{(0+2+4)}$ (solid),
together with the simulation data (markers with $1\sigma$ error
bars). The lower panels display the residual
$\xi_{\mathrm{pert}} - \xi_{\mathrm{sim}}$ with a shaded band
indicating the combined simulation uncertainty. For both $\xi_{11}$
and $\xi_{22}$, the fourth-order prediction falls within the noise
envelope across the full time range.

\begin{figure*}[t]
  \centering
  \includegraphics[width=\textwidth]{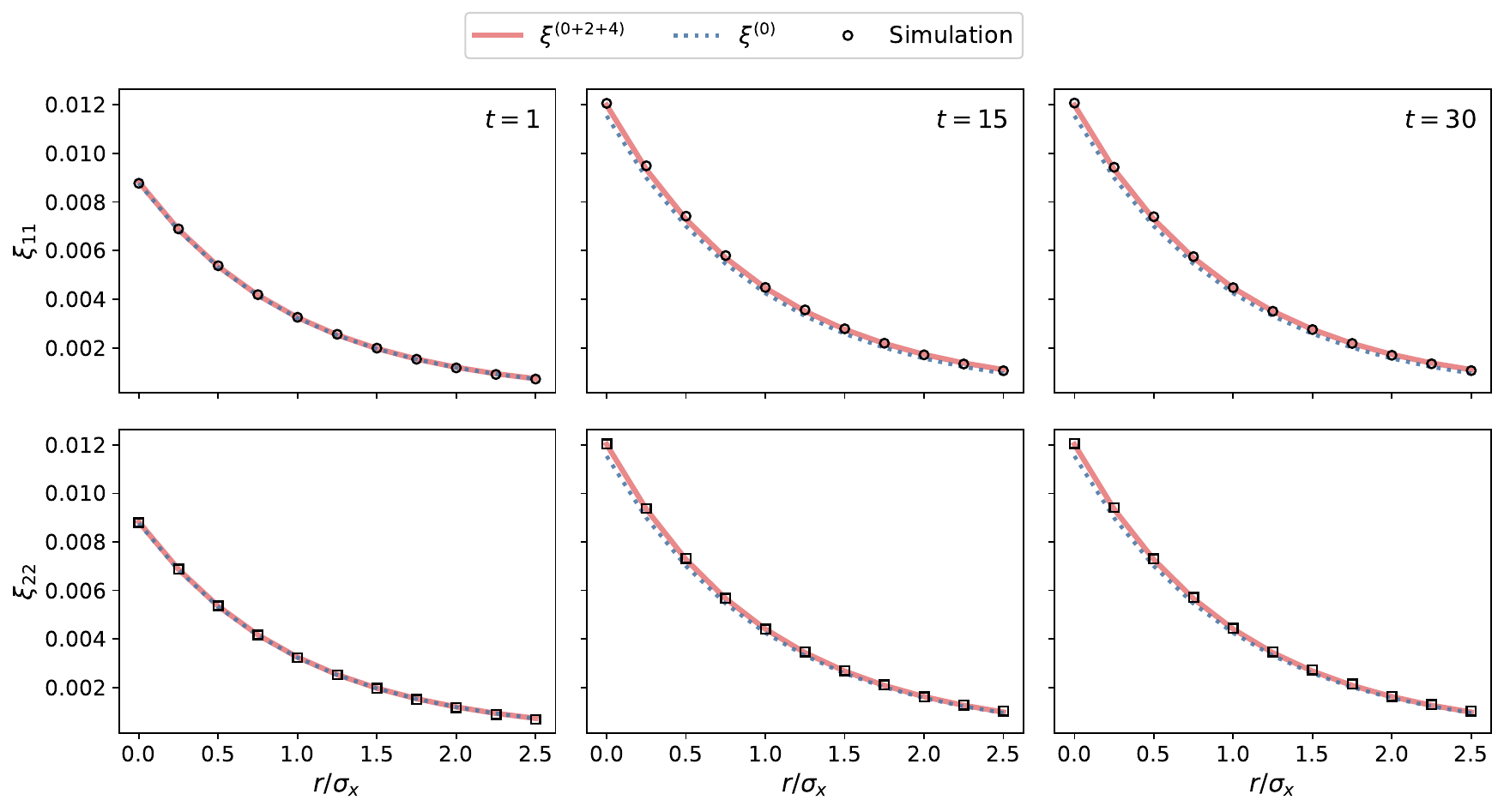}
  \caption{Correlators versus spatial separation at three times.
    Dotted: $\xi^{(0)}$; solid: $\xi^{(0+2+4)}$;
    markers: simulation.
    Rows: $\xi_{11}$ and $\xi_{22}$;
    columns: $t = 1,\;15,\;30$; simulation markers at the nearest cached time,
    $33.6$ in the last column.}
  \label{fig:xi-vs-r}
\end{figure*}

\Cref{fig:xi-vs-r} presents the spatial dependence at three
measurement times ($t = 1$, $15$, $30$), spanning the transient
build-up and the approach to stationarity (the linear relaxation time
is $1/\gamma = 1$). The zeroth-order free theory already captures the
overall shape; the fourth-order correction brings the prediction into
quantitative agreement with the simulation at all separations.

\begin{figure*}[t]
  \centering
  \includegraphics[width=0.95\textwidth]{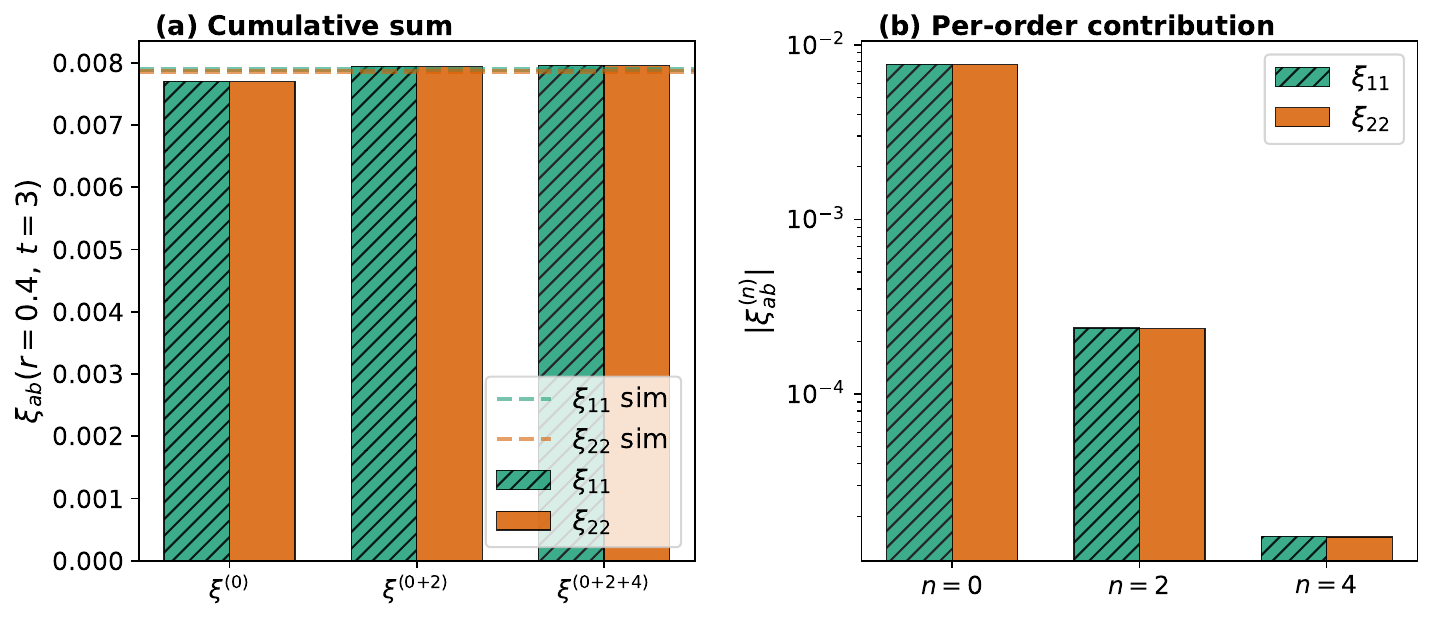}
  \caption{Order-by-order contributions at $r = 0.4\,\sigma_x$, $t = 3$.
    (a)~Cumulative sum for $\xi_{11}$ (hatched bars) and $\xi_{22}$
    (solid bars); dashed lines mark the simulation values at the nearest
    cached time, $t = 2.64$.
    (b)~Per-order contributions $|\xi^{(n)}|$, showing geometric
    suppression $\sim\!\lambda^{n/2}$.}
  \label{fig:convergence}
\end{figure*}

The suppression of successive orders is illustrated in
\cref{fig:convergence} at the representative point
$r = 0.4\,\sigma_x$, $t = 3$. Each successive even-order
contribution is suppressed by $\sim\!\lambda \approx 0.05$ relative
to the previous one over the computed range, so the expansion parameter
is small and the fourth-order truncation is adequate at this coupling;
whether the suppression persists at higher orders is not established,
and the series may be asymptotic. The computed corrections are all
positive (a consequence of the MSR sign factors and the positivity of
the coupling tensor), so each partial sum lies below the next.

The residual between perturbative and simulated correlators has four
contributions. The dominant one is \emph{simulation noise}
($\sim\!0.6\%$), the irreducible statistical uncertainty from the
finite ensemble. The \emph{truncation error} from omitting orders
$\geq 6$ is estimated at $\lesssim\!0.3\%$ from the observed
geometric suppression. The \emph{quadrature error} of the perturbative
side is below $0.1\%$ for $t \lesssim 15$, rising at the largest time
plotted, where both channels have long since saturated, to about $2\%$ of
the order-2 channel at the plotted separation ($3\%$ at the extreme of
the grid) and $2\%$ of the order-4 channel relative to its saturated
value; the plotted totals move by at most $0.1\%$. The \emph{time-stepping bias} of the Heun
integrator ($O(\Delta t^2) \approx 0.04\%$) is negligible by
comparison. In summary, the agreement between the perturbative
calculation and the simulation is limited by the statistical precision
of the simulation.

\section{Worked example II: non-Gaussian driving field and a non-local vertex}
\label{sec:example2}

The example of \cref{sec:examples} used Gaussian noise, so only the
two-point cumulant of the driving field entered and every diagram was
built from $\RespOp$ and $\CorrOp$ propagators alone. This section tests
the part of the formalism that handles non-Gaussian noise: the higher
cumulants of the driving field, which enter the MSR action of
\cref{sec:msr-action} as vertices whose legs sit at different points of
spacetime. We keep the same Langevin system and coupling and change only
the noise, from a Gaussian field to a train of random pulses. This
shot noise is chosen because every one of its cumulants is known in
closed form. At zero coupling the perturbative result can then be
compared with an exact answer, and at finite coupling the leading terms
left out of a lower-order expansion can be computed. A single number, the mean number of pulses per correlation
cell $\sigma_t\sigma_x$, sets how far the noise
is from Gaussian, so the non-Gaussian effect can be varied with
everything else held fixed. The observable is the cross-correlator
$\xi_{12}$, which vanishes for Gaussian noise and therefore measures the
non-Gaussian contribution directly. \Cref{sec:example2-diagrams} derives
the diagrams and the selection rule behind that observable,
\cref{sec:example2-evaluation} describes how the non-local vertex is
evaluated, and \cref{sec:example2-budget} gives the results and the
error budget.

Concretely, we keep the Langevin system and $\gamma = 1$ of
\cref{sec:examples}, scale its coupling tensor by an amplitude $s$,
$F^{(3)}_{abc} \to s\,F^{(3)}_{abc}$, and replace the Gaussian driving field by a filtered Poisson process: independently for
each component, events $(x_k,s_k)$ are drawn from a Poisson process of
rate $\nu$ over the line and the whole past, and
\begin{equation}
  \begin{gathered}
  \eta_a(x,t) = \sum_k h\,w(x-x_k)\,g(t-s_k) - \vev{\cdot},
  \\
  g(\tau) = \Theta(\tau)\,e^{-\tau/\sigma_t},
  \quad w(x) = e^{-|x|/\sigma_x}.
  \end{gathered}
  \label{eq:shot-noise}
\end{equation}
Campbell's theorem \cite{Rice1944} gives every cumulant in closed form, all factoring
through the single source point $(x',s)$,
\begin{equation}
  \begin{split}
  \cumulant^{(m)}_{a_1\ldots a_m}(\bm{z}_1\ldots\bm{z}_m)
    &= \delta_{a_1\ldots a_m}\,\nu h^{m} \\
    &\quad X_m(x_1\ldots x_m)\,T_m(t_1\ldots t_m),\\
  X_m &= \int\!\dd x'\prod_{i} w(x_i-x')\,,
  \end{split}
  \label{eq:campbell}
\end{equation}
with $T_m$ the analogous integral over the source time. We take
$\nu = 2$, $h = 1$, $\sigma_t = 0.5$ and $\sigma_x = 1$. The cumulants
are component-diagonal, as in \cref{sec:examples}, and the odd ones do
not vanish.

\begin{figure*}[tbp]
  \centering
  \includegraphics[width=\textwidth]{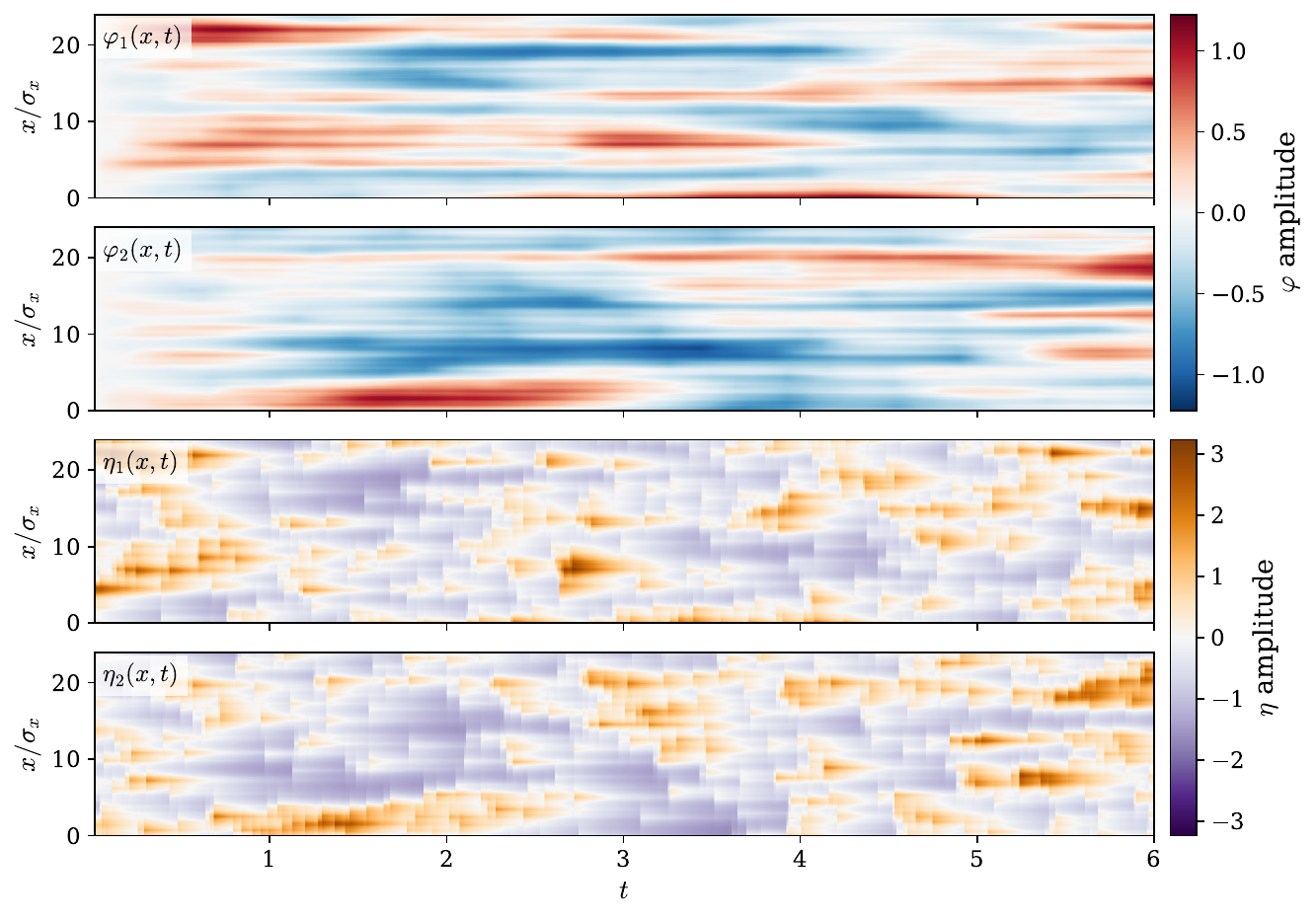}
  \caption{One realisation of the shot-noise example at $s = 0.2$. \emph{Top}:
    $\varphi_1$ and $\varphi_2$. \emph{Bottom}: the driving fields $\eta_1$ and
    $\eta_2$, in which each Poisson event appears as a sharp onset decaying in
    time and space. Separate colour scales for $\varphi$ and $\eta$, as in
    \cref{fig:waterfall}. The driving
    field is evaluated exactly from its events; the physical fields are
    integrated with the Heun scheme.}
  \label{fig:ex2-waterfall}
\end{figure*}

The shape of the one-point distribution depends only on
$n \equiv \nu\sigma_t\sigma_x$,
\begin{equation}
  \text{skewness} = \frac{4\sqrt2}{9}\,\frac{1}{\sqrt n},
  \qquad
  \text{excess kurtosis} = \frac{1}{2n},
  \label{eq:one-knob}
\end{equation}
with $n\to\infty$ the Gaussian limit. Since the variance
$\cumulant^{(2)}(\bm{z},\bm{z}) = nh^2/2$ can be held fixed by adjusting $h$, a sweep in $n$ changes the
non-Gaussian sector while leaving the Gaussian sector unchanged. The run below uses $n = 1$
(skewness $0.63$). \Cref{fig:ex2-waterfall} shows one realisation: each
event enters as a sharp onset decaying in time and space, so the driving
field is a superposition of discrete pulses rather than the smooth field
of \cref{fig:waterfall}.

\begin{figure}[tbp]
  \centering
  \includegraphics[width=\columnwidth]{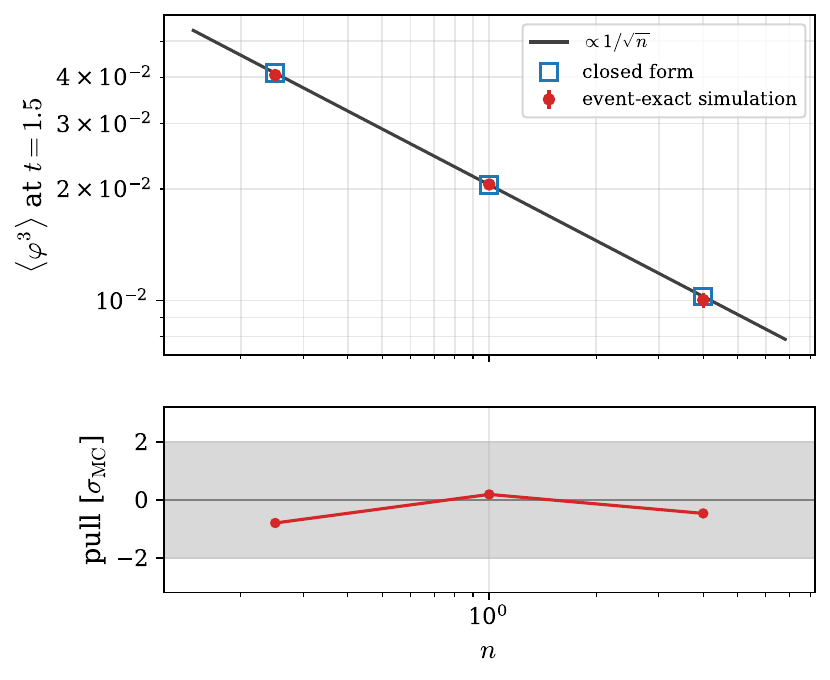}
  \caption{$\vev{\varphi^3}$ at $t = 1.5$ against $n = \nu\sigma_t\sigma_x$, with
    $\cumulant^{(2)}$ held fixed. Line: the $1/\sqrt n$ law of
    \cref{eq:one-knob}; squares: closed form; circles: simulation.
    \emph{Bottom}: pulls in units of the Monte Carlo error, $\pm2\sigma$ band
    shaded.}
  \label{fig:ex2-scaling}
\end{figure}

\subsection{Non-local vertices, diagram classes and selection rules}
\label{sec:example2-diagrams}

In the MSR action of \cref{sec:msr-action} the non-Gaussian part of the
cumulant generating functional contributes response-field-only vertices,
\begin{multline}
  \Sint^{\mathrm{K}} =
  \frac{\ii}{3!}\int\!\dd\bm{z}_1\,\dd\bm{z}_2\,\dd\bm{z}_3\;
  \cumulant^{(3)}_{abc}(\bm{z}_1,\bm{z}_2,\bm{z}_3) \\
  \psi_a(\bm{z}_1)\,\psi_b(\bm{z}_2)\,\psi_c(\bm{z}_3)
  - \frac{1}{4!}\int\!\dd\bm{z}_1\cdots\dd\bm{z}_4 \\
  \cumulant^{(4)}_{abcd}(\bm{z}_1,\ldots,\bm{z}_4)\,
  \psi_a(\bm{z}_1)\cdots\psi_d(\bm{z}_4) + \cdots\,,
  \label{eq:Sint-K}
\end{multline}
which are non-local: the $\psi$-legs sit at distinct spacetime points and
the coupling is a kernel rather than a tensor. With the local cubic
vertex $\Sint^{\mathrm{F}}$ of \cref{sec:examples},
$\Sint = \Sint^{\mathrm{F}} + \Sint^{\mathrm{K}}$. In \sftwick{} the
cumulant vertices are \texttt{NonLocalVertex} objects of the workflow
layer whose coupling is a callable evaluating \cref{eq:campbell}; contraction, canonicalisation and
grouping proceed as for local vertices.

Which cumulant enters at which order follows from counting legs. A
diagram for $\langle\varphi_a\varphi_b\rangle$ with $n_F$ cubic vertices
and one cumulant vertex of order $m$ contains $2 + 2n_F$ physical and
$n_F + m$ response operators; each $\psi$ must pair with a $\varphi$ and
the remaining $\varphi$'s pair among themselves, so
\begin{equation}
  2 + n_F - m \;\geq\; 0 \quad\text{and even}.
  \label{eq:leg-count}
\end{equation}
Hence $\cumulant^{(m)}$ first enters at order $m - 1$: $\cumulant^{(3)}$
at order 2 (the FK class), $\cumulant^{(4)}$ at order 3 (FFK4) and
$\cumulant^{(5)}$ at order 4.

At second order $\Sint^2$ produces the six FF diagrams of
\cref{fig:example-diagrams} and the two FK diagrams of
\cref{fig:example2-FK-diagrams}; the latter carry four
$\RespOp$-propagators and no $\CorrOp$. With $F_{122} : F_{212} : F_{221} = 1 : \tfrac12 : \tfrac12$ and a
component-diagonal cumulant, the FK coupling sum collapses to
$F_{abb}\,\cumulant^{(3)}_{bbb}$ and $F_{baa}\,\cumulant^{(3)}_{aaa}$, so
\begin{equation}
  \xi^{\mathrm{FK}}_{11} = \xi^{\mathrm{FK}}_{22} = 0,
  \qquad
  \xi^{\mathrm{FK}}_{12} \neq 0\,.
  \label{eq:FK-selection}
\end{equation}
The leading non-Gaussian signal therefore appears in the cross-correlator
$\xi_{12}$, which vanishes identically in the Gaussian example. The
reason is a symmetry: the Gaussian theory with this $F$ is invariant
under $(\varphi_2, \psi_2) \to (-\varphi_2, -\psi_2)$. Every vertex with
an even number of index-2 fields, i.e.\ $\Sint^{\mathrm{F}}$ and the even
cumulants, respects it; only the odd cumulants break it. Consequently
order 0, FF, FFFF and FFK4 do not contribute to $\xi_{12}$: through third
order it is the FK channel alone, followed at fourth order by
$F^{3}\cumulant^{(3)}$ and $F^{3}\cumulant^{(5)}$. The diagonal
correlators receive order 0, FF, FFK4 and FFFF, but no FK.

\begin{figure*}[tbp]
  \centering
  \tikzset{fkext/.style={circle, inner sep=0pt, minimum size=6.5pt, fill=black},
           fkvert/.style={rectangle, inner sep=0pt, minimum size=8pt, fill=black},
           fkR/.style={draw={rgb,255:red,214;green,96;blue,77}, dashed, line width=1.6pt, -{Latex[length=2mm]}},
           fklab/.style={font=\small}}
  \begin{tabular}{@{}c@{\hspace{0.6cm}}c@{}}
  \begin{tikzpicture}[scale=0.68, every node/.style={inner sep=0pt}, baseline=(current bounding box.center)]
    \node[fkext, label={[label distance=3pt,fklab]left:$\varphi_{a}(\bm{x})$}] (ext0) at (-2.8, 0) {};
    \node[fkext, label={[label distance=3pt,fklab]right:$\varphi_{b}(\bm{y})$}] (ext1) at (2.8, 0) {};
    \node[fkvert, label={[label distance=2.5pt,fklab]above left:$F\,(\bm{z}_0)$}] (F) at (-0.84, 0.44) {};
    \node[fkvert, label={[label distance=2.5pt,fklab]below right:$K\,(\bm{z}_1,\bm{z}_2,\bm{z}_3)$}] (K) at (0.84, -0.44) {};
    \draw[fkR] (F) -- (ext0);
    \draw[fkR] (K) -- (ext1);
    \draw[fkR] (K) to[bend left=28] (F);
    \draw[fkR] (K) to[bend right=28] (F);
  \end{tikzpicture}
  &
  \begin{tikzpicture}[scale=0.68, every node/.style={inner sep=0pt}, baseline=(current bounding box.center)]
    \node[fkext, label={[label distance=3pt,fklab]left:$\varphi_{a}(\bm{x})$}] (ext0) at (-2.8, 0) {};
    \node[fkext, label={[label distance=3pt,fklab]right:$\varphi_{b}(\bm{y})$}] (ext1) at (2.8, 0) {};
    \node[fkvert, label={[label distance=2.5pt,fklab]above right:$F\,(\bm{z}_0)$}] (F) at (0.84, 0.44) {};
    \node[fkvert, label={[label distance=2.5pt,fklab]below left:$K\,(\bm{z}_1,\bm{z}_2,\bm{z}_3)$}] (K) at (-0.84, -0.44) {};
    \draw[fkR] (K) -- (ext0);
    \draw[fkR] (F) -- (ext1);
    \draw[fkR] (K) to[bend left=28] (F);
    \draw[fkR] (K) to[bend right=28] (F);
  \end{tikzpicture}
  \\[2pt]
  {\footnotesize (a)\ $\RespOp(\bm{x};\bm{z}_0)\,\RespOp(\bm{y};\bm{z}_1)\,\RespOp(\bm{z}_0;\bm{z}_2)\,\RespOp(\bm{z}_0;\bm{z}_3)$} &
  {\footnotesize (b)\ $\RespOp(\bm{y};\bm{z}_0)\,\RespOp(\bm{x};\bm{z}_1)\,\RespOp(\bm{z}_0;\bm{z}_2)\,\RespOp(\bm{z}_0;\bm{z}_3)$}
  \end{tabular}
  \caption{The two order-two FK diagrams of
    $\langle\varphi_a(\bm{x})\,\varphi_b(\bm{y})\rangle$. The local vertex $F$
    sits at $\bm{z}_0$; the non-local vertex $K$ has its three $\psi$-legs at
    $\bm{z}_1, \bm{z}_2, \bm{z}_3$. All four propagators are $\RespOp$ (red
    dashed, arrow from $\psi$ to $\varphi$). Diagram~(b) is (a) with
    $\bm{x} \leftrightarrow \bm{y}$.}
  \label{fig:example2-FK-diagrams}
\end{figure*}

\subsection{Contracting the legs of the cumulant vertex}
\label{sec:example2-evaluation}

The spatial delta functions of the $\RespOp$-propagators collapse the FK
integrals as in \cref{sec:examples}, leaving a four-dimensional integral
over the vertex times. When the cumulant kernel is narrow in the relative
times compared with the outer range, a fixed tensor-product rule on this
integral does not converge; the deformed-Gaussian example shipped as
\texttt{examples/demo2} is such a case. \sftwick{} therefore allows the
leg integrals to be performed inside the kernel, through the
\texttt{already\_R\_contracted} flag of a non-local vertex: the user
callable returns
\begin{multline}
  K_R(t_1', \ldots, t_m';\,\bm{x}')
  = \int\!\dd u_1\cdots\dd u_m \\
    \prod_{i=1}^{m} R(t_i', u_i)\;
    \cumulant^{(m)}(u_1, \ldots, u_m;\,\bm{x}')\,,
  \label{eq:R-contracted}
\end{multline}
where $t_i'$ are the times of the partners of the $m$ $\psi$-legs, and
the outer integral of an FK diagram becomes one-dimensional.

For the shot noise \cref{eq:shot-noise} this contraction is exact: every
cumulant factors through one source point, so the $m$ leg integrals
collapse to a single integral over the source time with a $2^m$-term
closed form. The correlation propagator uses the temporal closed form of
\cref{app:propagators} with the spatial envelope $X_2$ supplied by a short
module, so no $\CorrOp$ quadrature runs. The same code gives $K_R$ at any order
$m$, so the $\cumulant^{(5)}$ term below can be computed.

\subsection{Results and error budget}
\label{sec:example2-budget}

\begin{figure*}[tbp]
  \centering
  \includegraphics[width=\textwidth]{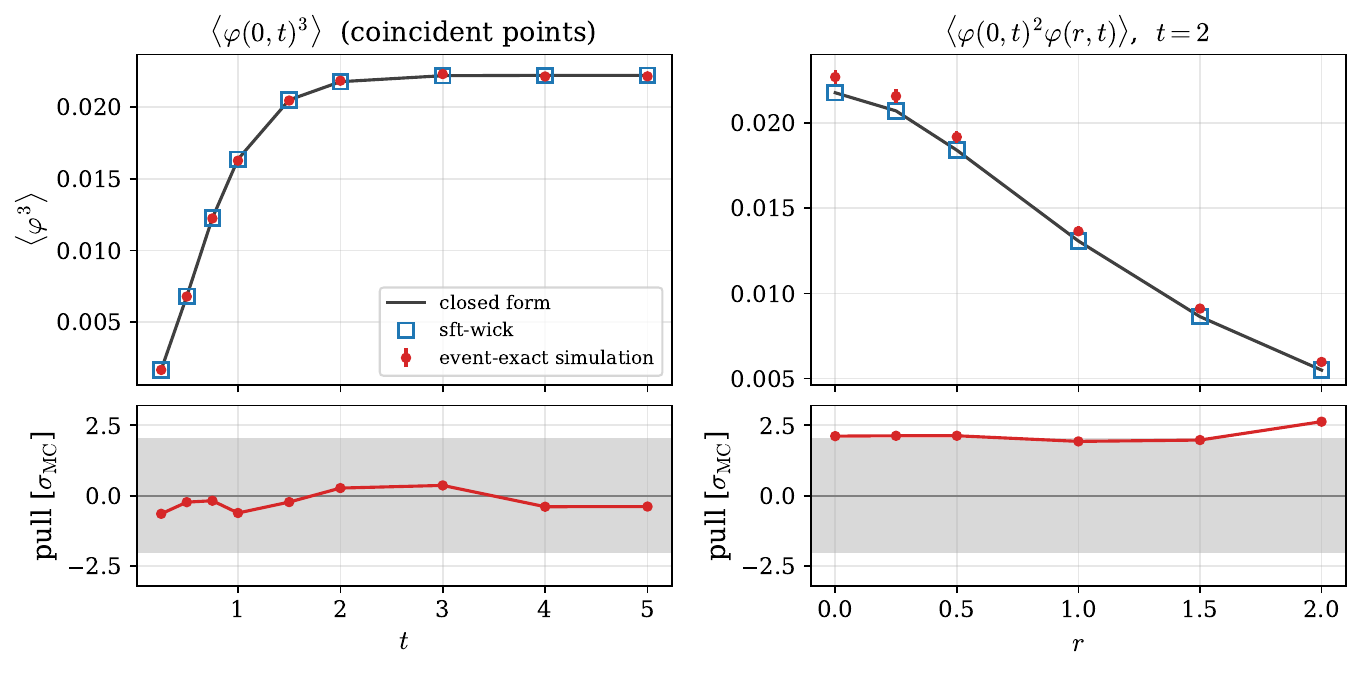}
  \caption{The free-field limit $F = 0$. \emph{Left}: $\vev{\varphi(0,t)^3}$ against
    time. \emph{Right}: $\vev{\varphi(0,t)^2\varphi(r,t)}$ against separation
    at $t = 2$. Line: closed form; open squares: \sftwick{}; circles: the
    event-exact simulation. \emph{Bottom}: pulls, $\pm2\sigma$ band shaded.
    Points within a panel share realisations, so the offset in the right-hand
    panel is a single fluctuation.}
  \label{fig:ex2-level-a}
\end{figure*}

At $F = 0$ the series terminates: the connected $m$-point function is a
single diagram equal to the closed form of \cref{eq:R-contracted}.
\sftwick{} reproduces it to $1.4\times10^{-16}$ relative at $m = 3$ and
$2.0\times10^{-16}$ in the separation sweep (\cref{fig:ex2-level-a}), and
to $6.6\times10^{-16}$ at $m = 4$ (not plotted). The reference
simulation at $F = 0$ has no discretisation error: the drift has no
spatial derivative, so a finite set of sites is exact; the sites sit at
the plotted separations; and the response to an exponential pulse is
analytic per event, so there is no time stepping. The only approximation,
the finite event window, is bounded at $7.6\times10^{-11}$. Over
$1.2\times10^{6}$ realisations the pulls are $\leq0.64\sigma$ for
$\vev{\varphi^3}(t)$ and $\leq1.12\sigma$ for the connected
$\vev{\varphi^4}(t)$ (the latter not plotted); the common $2\sigma$ offset
of the separation panel of \cref{fig:ex2-level-a} is a single fluctuation
shared by its points. The same limit tests \cref{eq:one-knob}:
\cref{fig:ex2-scaling} sweeps $n = 0.25, 1, 4$ at fixed
$\cumulant^{(2)}$, and the simulation reproduces the predicted factor of
four as $4.04$, within its Monte Carlo error.

\begin{figure*}[tbp]
  \centering
  \includegraphics[width=\textwidth]{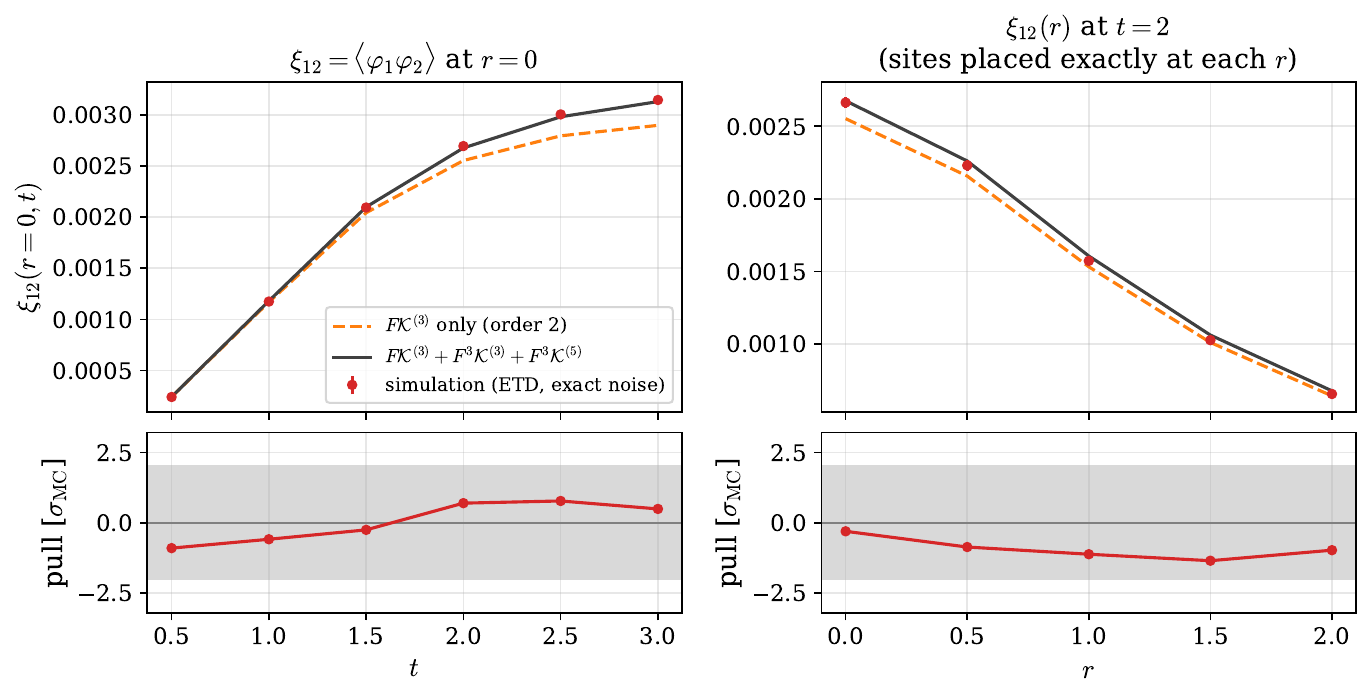}
  \caption{$\xi_{12}$ at $s = 0.2$. \emph{Left}: against time at $r = 0$;
    \emph{right}: against separation at $t = 2$. Dashed: the $F\cumulant^{(3)}$
    channel alone; solid: all three computed channels; circles: simulation.
    \emph{Bottom}: pulls, $\pm2\sigma$ band shaded.}
  \label{fig:ex2-xi12}
\end{figure*}

\begin{figure*}[tbp]
  \centering
  \includegraphics[width=\textwidth]{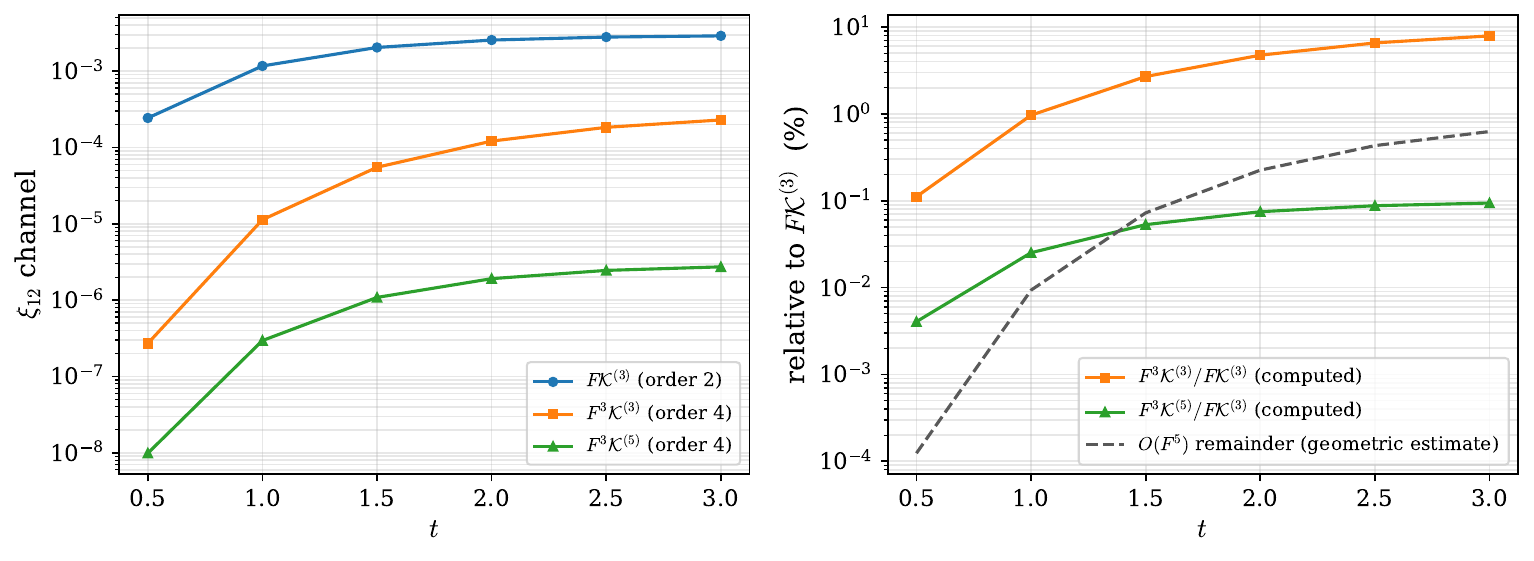}
  \caption{The channels of \cref{tab:ex2-budget} against time. \emph{Left}: the
    three computed channels. \emph{Right}: the two corrections as a percentage
    of $F\cumulant^{(3)}$, with the geometric estimate
    $(F^{3}\cumulant^{(3)}/F\cumulant^{(3)})^2$ of the uncomputed $O(F^{5})$
    remainder.}
  \label{fig:ex2-channels}
\end{figure*}

With $F$ switched on at amplitude $s = 0.2$, all three leading channels
of $\xi_{12}$ are computed: the order-2 $F\cumulant^{(3)}$, the order-4
$F^{3}\cumulant^{(3)}$ (30 diagrams) and $F^{3}\cumulant^{(5)}$
(6 diagrams). At $t = 3$ the two corrections are $7.9\%$ and $0.094\%$ of
the leading channel (\cref{fig:ex2-channels}), leaving an $O(F^5)$
remainder, estimated by geometric extrapolation at $0.63\%$, as the only
uncomputed term (\cref{tab:ex2-budget}). Against an exponential-time-differencing
\cite{CoxMatthews2002} simulation (\cref{fig:ex2-xi12}; six seeds for
$\xi_{12}(t)$, one for the other two comparisons), $\xi_{12}(t)$ agrees
within $0.90\sigma$ at every time, $\xi_{12}(r)$ within $1.35\sigma$ and
the diagonal $\xi_{11}$, computed through its FF and FFK4 classes (the order-4 FFFF
class lies below the Monte Carlo error and is omitted), within $0.69\sigma$; none of the
$7.3\times10^{6}$ trajectories diverged. Since $F^{3}\cumulant^{(3)}$ and
$F^{3}\cumulant^{(5)}$ both scale as $s^3$, no amplitude-scaling test can
separate them; where a kernel has no closed form, the order of a residual
can instead be measured by scaling the coupling, as
\texttt{examples/demo2} does.

In summary, the non-local vertices of \cref{eq:Sint-K} are handled by the
same contraction engine as local ones, the leg-counting and selection
rules \cref{eq:leg-count,eq:FK-selection} are read off the reported
coupling sums, and the $R$-contracted vertex \cref{eq:R-contracted}
reduces a four-dimensional quadrature to a one-dimensional one. Not
established: the finite-$F$ comparison is a truncated series against a
discretised simulation, the $O(F^5)$ remainder is estimated rather than
computed, and \cref{eq:one-knob} is exhibited at three values of $n$
rather than fitted. The full set of checks and numbers ships with the
package (\texttt{examples/demo3/INTERPRETATION.md}); the demonstration
reproduces in about twelve minutes on 28 workers, or twenty on one.

\begin{table*}[tbp]
\centering
\caption{Error budget for $\xi_{12}(r{=}0,t)$ at $s=0.2$, in units of $10^{-3}$:
the three computed channels, their sum $\Sigma$, the pull of the simulation
against $\Sigma$, and the geometric estimate of the uncomputed $O(F^{5})$
remainder. Simulation: six seeds of $4\times10^{5}$ realisations; $\sigma$ is
the Monte Carlo error.}
\label{tab:ex2-budget}
\small
\begin{tabular}{@{}rrrrrrrr@{}}
\toprule
$t$ & $\xi_{12}^{\rm sim}\pm\sigma$ & $F\cumulant^{(3)}$ & $F^{3}\cumulant^{(3)}$ & $F^{3}\cumulant^{(5)}$ & $\Sigma$ & pull ($\sigma$) & $O(F^{5})$ est. \\
\midrule
  0.5 & $0.242\pm0.002$ & 0.243 & 0.0003 & 0.00001 & 0.244 & $-0.90$ & 0.00012\% \\
  1 & $1.175\pm0.010$ & 1.169 & 0.0113 & 0.00029 & 1.181 & $-0.58$ & 0.0093\% \\
  1.5 & $2.094\pm0.020$ & 2.042 & 0.0550 & 0.00108 & 2.099 & $-0.25$ & 0.073\% \\
  2 & $2.694\pm0.026$ & 2.553 & 0.1211 & 0.00191 & 2.676 & $+0.70$ & 0.23\% \\
  2.5 & $3.004\pm0.030$ & 2.794 & 0.1835 & 0.00245 & 2.980 & $+0.78$ & 0.43\% \\
  3 & $3.144\pm0.032$ & 2.897 & 0.2293 & 0.00273 & 3.129 & $+0.49$ & 0.63\% \\
\bottomrule
\end{tabular}
\end{table*}

\section{Algorithm}
\label{sec:algorithm}

Given an action $S = S_0 + \Sint$ and an observable
$\mathcal{O} = \varphi_{a_1}(\bm{x}_1)\cdots\varphi_{a_m}(\bm{x}_m)$,
\sftwick{} evaluates the perturbative expansion
$\vev{\mathcal{O}}_S = \sum_n (-1)^n / n!\;
\vevfree{\mathcal{O}\,\Sint^n}$
through five stages.

\paragraph{1.\ Action expansion}
The interaction $\Sint = \sum_\alpha V_\alpha$ may contain multiple
vertex types.  At order $n$ the multinomial expansion of $\Sint^n$
generates all combinations of vertex copies; each copy is instantiated
with fresh component indices and spatial variables via a monotonic
counter.

\paragraph{2.\ Wick contraction}
\label{sec:contraction}
The free-theory expectation $\vevfree{\cdot}$ is evaluated by Wick's
theorem.  Rather than enumerating all $(2n{-}1)!!$ operator pairings
(which is intractable for multi-component fields beyond second order),
\sftwick{} operates at the \emph{spatial topology} level.  Operators
are grouped by spatial point and field type; valid $R$-assignments
(each $\psi$ paired with a $\varphi$ at some point) are enumerated
with three pruning rules applied during construction:
$\psi$--$\psi$ pairings are skipped,
equal-point $R$-pairings are forbidden under the \ito{} prescription,
and assignments closing a directed $R$-cycle are rejected via a
memoised depth-first search.
The remaining $\varphi$-operators at each point are paired into
$C$-propagators.  Each resulting \emph{spatial topology} represents
many operator-level pairings whose multiplicity is given analytically by
\begin{equation}
  \label{eq:multiplicity}
  \mu = \frac{\displaystyle\prod_{v} m_v!}
             {\displaystyle 2^{N_{\mathrm{self}}}
              \cdot \prod_{e} k_e!}\,,
\end{equation}
where $m_v$ counts the $\varphi$-operators available for $C$-pairing
at vertex~$v$, $N_{\mathrm{self}}$ is the number of $C$-self-loops,
and $k_e$ is the multiplicity of each distinct $C$-edge.
Within each topology, the component-index routings are recovered by
permuting operators among same-type edge slots and deduplicating.

\paragraph{3.\ Diagram collection}
\label{sec:simplification}
Topologies related by relabelling of integration variables are
identified and merged.  The default backend computes a canonical edge
list by trying all $k!$ permutations of the $k$ internal spatial
variables ($k\!\leq\!4$ in practice); an optional \textsc{pynauty}
\cite{Nauty2014} backend provides efficient graph canonicalisation
for higher orders.  Routings mapping to the same
canonical diagram are collected into a single \texttt{DiagramTerm}
whose \emph{coupling sum} encodes the symmetry of the coupling tensor
(e.g.\ $F_{abc} + F_{bac}$).  A union-find pass then simplifies
propagator indices under user-specified diagonal or isotropic
constraints ($R_{ab} \propto \delta_{ab}$,
$C_{ab} \propto \delta_{ab}$).

\paragraph{4.\ Coupling evaluation}
Each \texttt{DiagramTerm} partitions its summation indices into
\emph{propagator indices} (appearing in at least one propagator) and
\emph{coupling-only indices} (internal to the coupling sum).  The
method \texttt{evaluate\_coupling()} substitutes the user-supplied
NumPy coupling tensors, sums over the coupling-only indices, and
applies the rational prefactor $(-1)^n/n!$ together with the MSR
phase $(-\ii)^{n_R}$.  The result is a numerical array indexed by
the propagator indices, computed once per diagram.

\paragraph{5.\ Numerical integration}
\label{sec:evaluation}
The spatial analysis stage decomposes each diagram's abstract spatial
points into direction equivalence classes (from $R$-propagator
$\delta$-functions) and a time-ordering DAG (from $R$ causality).
The correlation propagator $\CorrOp$ is evaluated from a built-in
closed form for the family of diagonal constant drift with separable
noise whose temporal kernel is exponential, from a user-supplied closed-form
expression, or via double numerical integration of
$R \cdot \cumulant^{(2)} \cdot R$ with a node count fixed by a convergence
check (\cref{app:propagators}). Non-local cumulant vertices may be
supplied either as a raw kernel $\cumulant^{(m)}$ or, when the kernel is
narrow compared with the outer integration range, in $R$-contracted
form with the $\psi$-leg integrals performed inside the user callable
(\cref{sec:example2-evaluation}).  The time integrals
are performed with tensor-product Gauss--Legendre quadrature on the
causally ordered simplex, recommended for $d\!\leq\!5$ integration
variables, or with quasi-Monte Carlo sampling using Sobol
low-discrepancy sequences for higher-dimensional integrals; the latter is
the default and the former is selected explicitly.
Independent diagrams may be evaluated in parallel via \texttt{joblib}.

\begin{figure*}[tbp]
  \centering
  \includegraphics[width=\textwidth]{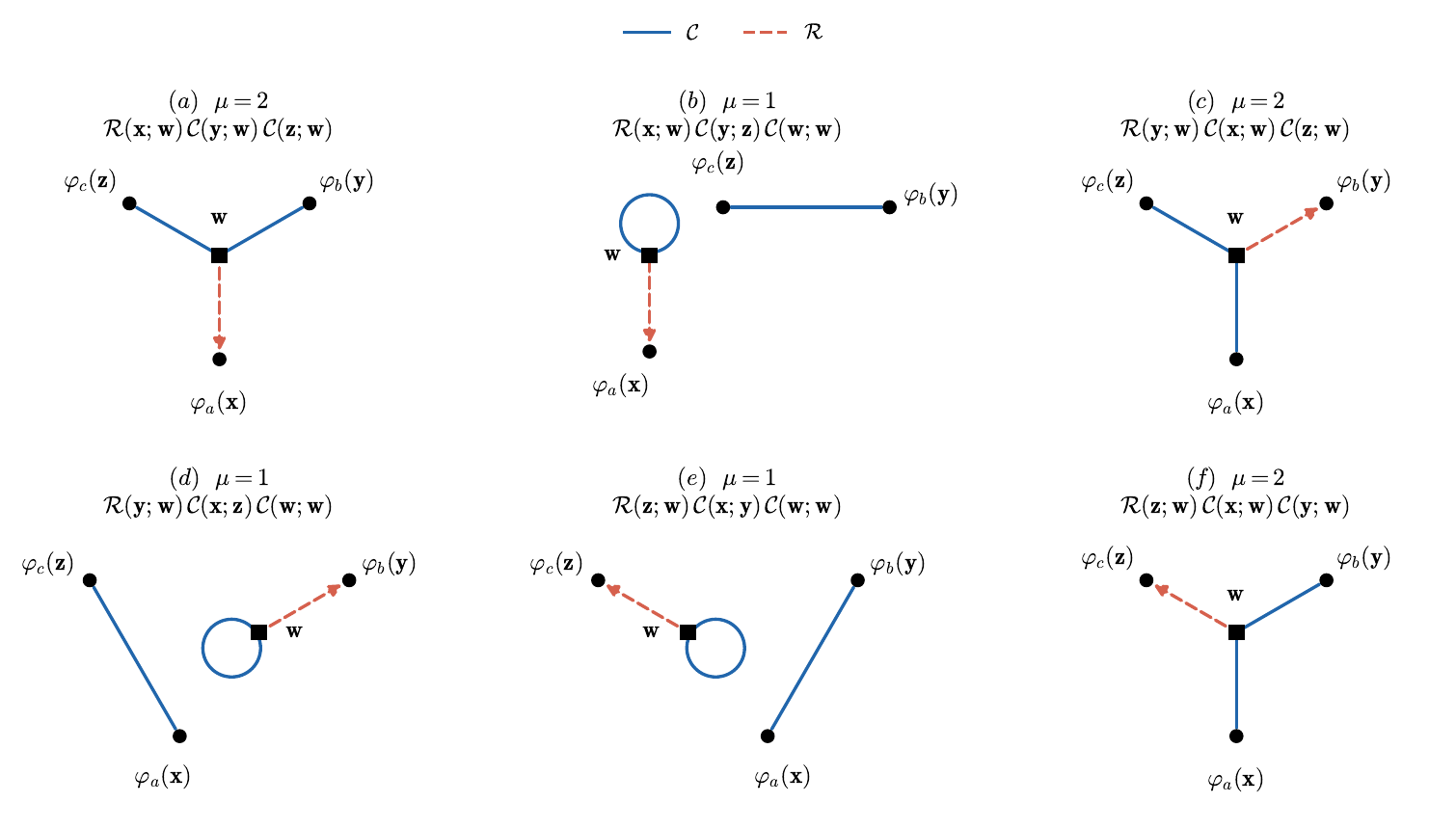}
  \caption{The six diagrams of the first row of \cref{tab:scaling},
    $\langle\varphi_a(\bm{x})\,\varphi_b(\bm{y})\,\varphi_c(\bm{z})\rangle$ at
    first order, as drawn by \sftwick{}; square: the cubic vertex $F$ at
    $\bm{w}$;
    circles: external points. Of the fifteen Wick pairings, six vanish by
    \cref{eq:ito} and the remaining nine group into these six topologies with
    multiplicities $\mu$. Blue solid: $\CorrOp$; red dashed with arrow:
    $\RespOp$.}
  \label{fig:order1-diagrams}
\end{figure*}

\paragraph{Complexity}
Direct Wick enumeration over the $3n{+}m$ operators of
$\Sint^n\,\mathcal{O}$, three per cubic vertex plus the $m$ external
operators, involves $(3n{+}m{-}1)!!$ pairings, factorial in the
perturbative order.  The topology engine replaces this by an
enumeration over spatial points (at most $n{+}m$) with analytic
multiplicities, whose count grows far more slowly (\cref{tab:scaling});
canonical-form
deduplication then collapses isomorphic topologies, and the remaining
component-index sums are absorbed analytically inside each
\texttt{DiagramTerm}.  \Cref{tab:scaling} reports the resulting
counts against $(3n{+}m{-}1)!!$ for a representative cubic vertex; the
observable alternates between a three-point and a two-point function
with the parity of $n$ so that the operator count is even.
\Cref{fig:order1-diagrams} shows the six diagrams behind its first row.
The three external operators carry distinct spatial labels; an
equal-point correlator is obtained by evaluating them at coincident
positions, which does not change the count.

\begin{table*}[t]
\centering
\caption{Scaling of the topology engine against direct Wick enumeration for the
cubic vertex $F_{abc}\,\psi_a\varphi_b\varphi_c$ with $N=3$: operator count,
raw Wick pairings, distinct diagrams returned, and their ratio. Wall-clock is
the minimum over four runs, each the best of three, on one core of an Apple
M3 Ultra. The first row's diagrams
are drawn in \cref{fig:order1-diagrams}.}
\label{tab:scaling}
\footnotesize\setlength{\tabcolsep}{3.5pt}
\begin{tabular}{@{}clccccc@{}}
\toprule
$n$ & Observable & Operators & Pairings & Diagrams & Wall-clock (s) & Reduction \\
 & & $3n{+}m$ & $(3n{+}m{-}1)!!$ & \sftwick{} & \sftwick{} & factor \\
\midrule
  1 & $\langle\varphi_a(x)\varphi_b(y)\varphi_c(z)\rangle$ & 6 & $15$ & 6 & $2.96\times10^{-4}$ & $2.5$ \\
  2 & $\langle\varphi_a(x)\varphi_b(y)\rangle$ & 8 & $105$ & 6 & $1.43\times10^{-3}$ & $17.5$ \\
  3 & $\langle\varphi_a(x)\varphi_b(y)\varphi_c(z)\rangle$ & 12 & $1.0\!\times\!10^{4}$ & 80 & $1.22\times10^{-1}$ & $130$ \\
\bottomrule
\end{tabular}
\end{table*}

The driver \texttt{compute\_moment}, called with an observable, an
action and an order, returns a \texttt{PerturbativeResult} containing
symbolic expressions,
\LaTeX{} rendering, and structured \texttt{DiagramTerm} objects ready
for numerical evaluation.

\section{Summary and outlook}
\label{sec:summary}

We have presented \sftwick{}, an open-source Python package that
automates perturbative calculations in stochastic field theories
formulated within the Martin--Siggia--Rose path-integral framework.
A user specifies an action and an observable; \sftwick{} returns the
perturbative expansion both as symbolic Feynman diagram expressions
(with full algebraic coefficients, ready for \LaTeX{} rendering) and
as numerical predictions obtained by integrating each diagram against
user-supplied propagator functions.

The central algorithmic idea is a \emph{spatial topology engine}
that enumerates topologically distinct diagrams, rather than
individual operator pairings, with analytically computed
multiplicities. Combined with MSR-specific pruning (vanishing
$\psi$--$\psi$ correlators, the \ito{} prescription, causal
$R$-loop elimination) and diagram isomorphism detection, this
avoids the full $(2n-1)!!$ enumeration of Wick pairings and makes
fourth-order and higher calculations tractable for multi-component
fields.

Validation against direct Langevin simulations of a two-component
system (\cref{sec:examples,sec:example2}) shows geometric suppression of
successive orders for Gaussian driving: at coupling
$\lambda = 0.05$, each successive even order is suppressed by
$\sim\!\lambda$, and the fourth-order prediction agrees with the
simulation within its statistical noise ($\lesssim\!0.6\%$). For a driving field
whose cumulants are non-vanishing at every order, the cross-correlator
that only odd cumulants can generate is reproduced by the
$F\cumulant^{(3)}$ channel at early times, and its late-time correction
is computed: both order-4 classes are evaluated, leaving an $O(F^5)$
remainder estimated at $0.63\%$. In the
free-field limit the package reproduces the closed form to machine
precision.

Several limitations remain. The canonical diagram form computation
scales as $k!$ in the number of integration variables; the optional
\textsc{pynauty} backend is recommended for orders $\geq 6$.
Higher-order driving-field cumulants $\cumulant^{(n)}$ with $n \geq 3$ are
supported by declaring non-local $\psi$-only vertices with a numeric or
callable kernel $\cumulant^{(n)}_{i_1\cdots i_n}(x_1,\dots,x_n)$ (no
additional contraction rules are needed, since such vertices enter
Wick's theorem on the same footing as any other interaction), and
\cref{sec:example2} shows that a kernel that is narrow in its relative
times should be supplied in $R$-contracted form.

Releases before 0.4.0 could not evaluate a diagram in which a cumulant
kernel's component index is carried onto a $\CorrOp$-propagator (the
order-4 $F^{3}\cumulant^{(3)}$ class among them), because the
callable-kernel path required those indices to be contracted away by the
$R$-propagators on its legs; since 0.4.0 they are evaluated, which is what
allows the leading omitted class of \cref{sec:example2-budget} to be
computed.

Looking ahead, planned extensions include integration with automatic differentiation
frameworks for gradient-based parameter inference, integration on
graphics processors for high-dimensional diagram integrals, momentum-space
representations for translationally invariant systems, and
resummation techniques to extend predictions beyond the weak-coupling
regime.

Finally, the statistical field theory framework for weak gravitational
lensing
that motivated the development of \sftwick{} is presented in a
companion paper \cite{Zhang2025_SFTLensing}, which applies the tools
described here to compute perturbative corrections to lensing
observables.


\section*{Acknowledgements}
The results were obtained as part of a project that has received
funding from the European Research Council (ERC) under the European
Union's Horizon 2020 research and innovation programme
(Grant agreement No.~948764).
The author also acknowledges support from the RadioForegroundsPlus
project HORIZON-CL4-2023-SPACE-01, GA 101135036.
The author thanks the two anonymous referees, whose reports prompted the
second worked example, the error budgets and the fast default path of
version~0.4.2 of the package. The author is indebted to the generations
of researchers whose published work forms the knowledge base on which this
paper and its software draw, and to the developers of the open-source
libraries and of the language-model technology used in building the
software.

\appendix
\crefalias{section}{appsec}\crefalias{subsection}{appsec}

\section{Installation, quick start and validation}
\label{app:installation}

This appendix describes version 0.4.2 of \sftwick{}, the version
archived with this paper. Wall-clock times are for one core of an Apple
M3 Ultra unless stated otherwise; ``laptop'' denotes the same runs on
the same machine throttled to background priority with a single thread,
intended as a proxy for a 2023 laptop, and ``CI''
denotes a stock GitHub \texttt{ubuntu-latest} runner.

\subsection{Installation}

\sftwick{} requires Python~3.10 or later:
\begin{lstlisting}[numbers=none]
pip install sft-wick               # this paper: sft-wick==0.4.2
pip install "sft-wick[progress]"   # optional: tqdm progress bars
\end{lstlisting}
The core dependencies are installed automatically: \texttt{numpy},
\texttt{scipy}, \texttt{networkx}, \texttt{matplotlib}, \texttt{pandas},
\texttt{pyyaml}, \texttt{tabulate} and \texttt{joblib}. Without
\texttt{tqdm}, progress is reported as plain lines on standard error. For a
development installation with the test and documentation extras:
\begin{lstlisting}[numbers=none]
git clone https://github.com/StatFieldTheory/sft-wick.git
cd sft-wick
pip install -e ".[dev,docs]"
\end{lstlisting}

\subsection{Quick start from the command line}
\label{app:quickstart}

The package exposes three layers: the low-level Python API of fields,
vertices and \texttt{compute\_moment} (\cref{app:api}); a workflow API
(\texttt{System}, \texttt{Expansion}, \texttt{Propagators}); and a YAML
configuration driven from the command line, which is the recommended
entry point. A one-command test of a fresh installation is
\begin{lstlisting}[numbers=none]
sft-wick quickstart
\end{lstlisting}
which writes the configuration below to the current directory and runs
it: the Gaussian model of \cref{sec:examples} at orders 0 and 2, three
separations, two times and $4096$ quasi-Monte Carlo samples, printed as
a Markdown table. It completes in 2~s (laptop proxy: 10~s; CI: 6~s).
\begin{lstlisting}[language={},numbers=none,basicstyle=\ttfamily\footnotesize]
system:
  field: {name: phi, n_components: 2}
  linear: {type: diagonal, gamma: [1.0, 1.0]}
  vertices:
    - name: F
      coupling:                 # bare F; the MSR factor is applied automatically
        - [[0.0, 0.0], [0.0, 1.0]]
        - [[0.0, 0.5], [0.5, 0.0]]
  noise:
    kappa2:
      type: separable_translation
      temporal: {type: exponential, lam: 0.05, sigma_t: 0.3}
      spatial:  {type: exponential, sigma_x: 1.0}
expand:
  observable: ["phi_a(x)", "phi_b(y)"]
  orders: [0, 2]
propagators:
  t_max: 5.0
  n_grid_t: 60
  c_closed_form: auto         # built-in closed form for this kernel family
sweep:
  positions_grid: {x: [0.0], y: [0.0, 0.5, 1.5]}
  t_final_grid: [1.0, 5.0]
  component_pairs: [[0, 0], [1, 1]]
  n_samples: 4096
  seed: 42
output:
  - {type: table, format: markdown}
\end{lstlisting}
Any configuration is run with \texttt{sft-wick run CONFIG};
\texttt{--override key=value} changes a parameter from the shell,
\texttt{--quiet} silences the progress output, and
\begin{lstlisting}[numbers=none]
sft-wick run CONFIG --dry-run
\end{lstlisting}
validates the file and prints a cost estimate before anything is
computed: the number of diagrams per order, of grid points and of
distinct separations, the propagator source that will be used, the
number of quadrature calls the chosen path will make, and a rough
wall-clock time from a one-second micro-benchmark of one propagator
call and one diagram evaluation per order. During a run, three stages
report progress with the elapsed time and an estimate of the remaining
time: the symbolic expansion, the build of the correlation-propagator
table, and the sweep over grid points and diagrams. The next steps up
from the quick start are \texttt{examples/demo1\_config.yaml} (orders
0--4, 71 diagrams; 50~s, laptop proxy 5.2~min, CI 1.5~min), the full sweep
behind the figures of \cref{sec:examples}
(\texttt{examples/demo1/L2/config.yaml}; about 15~min on 28 workers with
the Gauss--Legendre settings of \cref{sec:examples}) and the
configurations of \cref{sec:example2}
(\texttt{examples/demo3/config\_FK.yaml} and
\texttt{config\_F3K.yaml}). A continuous-integration job runs the quick start
and \texttt{demo1\_config.yaml} on every push to the main branch and
every pull request, under time limits of 600~s and 1200~s.

\subsection{Quick start from Python}

The same expansion is available from the low-level API, which is the
natural entry point for inspecting the diagram tables themselves:
\begin{lstlisting}
from sft_wick.fields import Field
from sft_wick.vertices import Vertex
from sft_wick.action import Action
from sft_wick.perturbation import compute_moment

# Define fields
phi = Field("phi", "physical", n_components=3)
psi = Field("psi", "response", n_components=3)

# Define cubic interaction vertex
vertex = Vertex(
    fields=[phi, phi, psi],
    coupling="F",
    local=True,
)

# Define action and observable
action = Action([vertex])
observable = [phi("a", "x"), phi("b", "y")]

# Compute to second order.  Order 1 vanishes identically here: two external
# phi plus one three-field vertex give five operators, which cannot be paired.
result = compute_moment(observable, action, order=2)

# Print LaTeX for each order
print(result.to_latex())

# Access structured diagram terms
for dt in result.diagram_terms(order=2):
    print(dt.to_latex())
\end{lstlisting}

\subsection{Propagator evaluation}
\label{app:propagators}

When no closed form applies, the cost of a run is dominated by the
double quadrature \cref{eq: Corr Op} that fills the $(t_1, t_2)$ table
of the correlation propagator. For a diagonal constant drift driven by
separable, translation-invariant noise with an exponential temporal
kernel (the family of both worked examples) the table is evaluated
from a closed form (\texttt{propagators.c\_closed\_form: auto}, the
default). The model of \cref{sec:example2} reaches it through
\texttt{c\_closed\_form\_module}, since its spatial envelope is not among
the kernels the YAML layer can name. For other kernels
\texttt{c\_method: auto} uses a Gauss--Legendre rule whose node count is
refined until the table's extreme cells converge to $10^{-8}$, falling
back to adaptive quadrature that splits the domain at the $|t_1 - t_2|$
cusp. The paths are validated against each other at their dispatch
boundaries, to $10^{-10}$ (closed form against Gauss--Legendre) and
$10^{-8}$ (against adaptive quadrature). Before release 0.3.0,
\texttt{examples/demo1\_config.yaml} with its closed-form hook disabled
(the configuration one referee ran) required $14\,400$ adaptive
double-quadrature calls at $74$--$245$~ms each, between 18 and 59 minutes
for the propagator table alone, and did not finish within the hour; the
present release makes none.

\subsection{Validation suite}
\label{app:validation}

The test suite of version 0.4.2 contains 1037 tests (parametrised cases
counted individually) in 38 files. What each file validates, against
which independent reference and at what tolerance, is recorded in a
generated catalogue, \texttt{docs/verification/catalog.rst}; a test
fails whenever the catalogue is out of date, so it cannot drift from the
code. \Cref{tab:validation} summarises it by subject. The references are
exact structural assertions and hand-derived expansions for the symbolic
engine; an independent brute-force Wick enumerator
(\texttt{tests/brute\_wick.py}) for multiplicities and canonical forms;
closed forms for the numerics; and, wherever a computation has an
alternative path, that path.

Because \texttt{pytest.approx(x, rel=\ldots)} compares against
$\max(\mathrm{rel}\times x,\,10^{-12})$, a tolerance written into a test
is not necessarily the one it enforces. The enforced tolerance is
therefore measured at every assertion site (\texttt{tools/approx\_\allowbreak audit.py})
and recorded in the catalogue. Of 207 runtime sites, 52 have a relative
tolerance in force, 153 pass an explicit absolute one and 2 compare
against zero; none is weakened by the floor, and a test refuses any new
site that would be. In earlier releases 22 sites were weakened and two
could not fail at all; in both vacuous cases the behaviour under test was
independently confirmed, so the suite's coverage had been overstated but
no result was wrong.

The sweep behind \cref{fig:xi-vs-time,fig:xi-vs-r,fig:convergence} is
re-run at each release, with fixed integration settings, as a regression
check on the library: the internal changes of 0.3.0 move no total by more
than $3.3\times10^{-14}$ over all 2016 entries, and those of 0.4.0 leave
every one bit-identical; 0.4.1 and 0.4.2 change nothing on this path. This
check says nothing about the deliberate change of integration settings
described in \cref{sec:examples}, which moves the plotted correlator by at
most $1.7\%$.

\begin{table*}[tbp]
\centering
\caption{The validation catalogue of \sftwick{} 0.4.2, summarised by
  subject; the generated catalogue lists every test file with what it
  checks, the reference it is checked against and the tolerance.}
\label{tab:validation}
\small
\begin{tabular}{@{}p{2.5cm}p{5.0cm}p{4.1cm}r@{}}
\toprule
Subject & What is checked & Reference & Tests \\
\midrule
Symbolic engine & expression tree, fields, Wick pairing rules, \texttt{compute\_moment} orders and coefficients, simplification & specification; hand-derived expansions; the two contraction engines against each other & 137 \\
Multiplicities and canonical forms & every pairing at order $\leq 2$ classified by its vanishing reason; topology multiplicities, \cref{eq:multiplicity}; canonical-form deduplication & brute-force Wick enumerator; isomorphic relabellings & 90 \\
Propagator numerics & closed-form, Gauss--Legendre and adaptive $\CorrOp$ tables at their dispatch boundaries; off-diagonal $\CorrOp$; time-dependent $\gamma(t)$; spline interpolation & each method against the others; closed-form OU propagators & 313 \\
Integrators & causal time simplex; $R$-contracted and callable $\cumulant^{(m)}$ vertices; cumulants of order 4 and 5; external response legs; all five backends & closed forms; brute-force Wick counting; backends against each other & 268 \\
Workflow and YAML & the \texttt{System}, \texttt{Expansion} and \texttt{Propagators} classes, YAML lowering, overrides, parallel layers & the low-level pipeline on the same physics & 68 \\
Drawing and \LaTeX{} & Matplotlib and TikZ renderers, \LaTeX{} output & specification; regression tests & 62 \\
Spectral propagators & disorder-averaged $R^{*}$, $C^{*}$ from a spectral density & OU closed form at a delta density & 57 \\
Self-consistency & fixed-point driver: convergence, divergence and oscillation reporting & analytic fixed points & 42 \\
\midrule
\multicolumn{3}{@{}l}{Total} & 1037 \\
\bottomrule
\end{tabular}
\end{table*}

\section{API reference}
\label{app:api}

This appendix collects the user-facing entry points of \sftwick{}.
A complete reference with all parameters and return types is included
in the repository documentation; here we summarise the principal
public API (\cref{tab:api}).

\begin{table*}[tbp]
\centering
\caption{Modules and principal public symbols in \sftwick{}.}
\label{tab:api}
\scriptsize\setlength{\tabcolsep}{3.5pt}
\begin{tabular}{@{}llp{5.6cm}@{}}
\toprule
\textbf{Module} & \textbf{Class/Function} & \textbf{Description} \\
\midrule
\multicolumn{3}{@{}l}{\textit{Field definitions}} \\
\texttt{fields} & \texttt{Field} & Declare a field type ($\varphi$ or $\psi$) with component count \\
                & \texttt{FieldOperator} & Concrete field instance with UID, index, spatial arg \\
\midrule
\multicolumn{3}{@{}l}{\textit{Interaction structure}} \\
\texttt{vertices} & \texttt{Vertex} & Interaction-term template (local or non-local) \\
                  & \texttt{VertexInstance} & Fresh copy with collision-free indices \\
\texttt{action}   & \texttt{Action} & Collection of vertices; multinomial expansion \\
\midrule
\multicolumn{3}{@{}l}{\textit{Contraction engines}} \\
\texttt{wick} & \texttt{wick\_contract()} & Operator-level Wick contraction \\
              & \texttt{wick\_contract\_spatial()} & Spatial-topology enumeration \\
\midrule
\multicolumn{3}{@{}l}{\textit{Main drivers}} \\
\texttt{perturbation} & \texttt{compute\_moment()} & Symbolic expansion pipeline \\
                      & \texttt{compute\_moment\_numerical()} & Fast path for high orders \\
                      & \texttt{PerturbativeResult} & Container for order-by-order results \\
                      & \texttt{DiagramTerm} & Structured diagram for numerical evaluation \\
\midrule
\multicolumn{3}{@{}l}{\textit{Simplification}} \\
\texttt{simplify} & \texttt{simplify()} & Multi-pass expression simplification \\
                  & \texttt{collect\_by\_diagram()} & Group by Feynman diagram isomorphism \\
                  & \texttt{diagonal\_propagators()} & Enforce diagonal/isotropic constraints \\
\midrule
\multicolumn{3}{@{}l}{\textit{Numerical evaluation}} \\
\texttt{evaluate} & \texttt{analyze\_spatial()} & Spatial structure analysis (DAG, groups) \\
                  & \texttt{PropagatorModel} & User-provided propagator functions \\
                  & \texttt{PropagatorCache} & Cached $C$ propagator with spline interpolation \\
                  & \texttt{DiagramIntegrand} & Ready-to-integrate object \\
                  & \texttt{integrate\_two\_point\_qmc()} & QMC integration for two-point functions \\
\midrule
\multicolumn{3}{@{}l}{\textit{Visualisation}} \\
\texttt{diagrams} & \texttt{FeynmanDiagram} & \textsc{NetworkX} multigraph representation \\
\texttt{drawing}  & \texttt{DiagramRenderer} & Matplotlib rendering \\
\texttt{latex}    & \texttt{LaTeXFormatter} & Configurable \LaTeX{} output \\
\bottomrule
\end{tabular}
\end{table*}

\paragraph{Declaring fields, vertices and actions}
The theory is specified through three lightweight constructors:
\begin{lstlisting}
Field(name: str,
      field_type: Literal["physical", "response"] | FieldType,
      n_components: int = 1)

Vertex(fields: Sequence[Field],
       coupling: str,
       local: bool = True,
       equal_time: bool = False,
       already_R_contracted: bool = False)

Action(vertices: list[Vertex])
\end{lstlisting}
A \texttt{Field} declares a physical ($\varphi$) or response ($\psi$)
field and its internal component count; a \texttt{Vertex} pairs an
ordered list of \texttt{Field}s with a coupling-tensor name and a
locality flag (set \texttt{local=False} for kernel-valued
non-local interactions); the \texttt{Action} is simply a list of
vertices that together generate $\Sint$.

\paragraph{Driving the perturbative pipeline}
The symbolic entry point is
\begin{lstlisting}
compute_moment(
    observable: list[FieldOperator],
    action: Action,
    order: int,
    ito: bool = True,
    response_phase: bool = True,
    collect_topology: bool = True,
    diag_R: bool = False,  diag_C: bool = False,
    iso_R:  bool = False,  iso_C:  bool = False,
) -> PerturbativeResult
\end{lstlisting}
which returns a \texttt{PerturbativeResult} whose
\texttt{diagram\_terms(order)} method exposes the list of
\texttt{DiagramTerm} objects at each perturbative order.

\paragraph{Numerical evaluation}
The method that turns a \texttt{DiagramTerm} into a callable
\texttt{DiagramIntegrand} object is
\begin{lstlisting}
term.build_integrand(
    coupling_values: dict[str, np.ndarray],
    fixed_indices: dict[str, int] | None = None,
) -> DiagramIntegrand
\end{lstlisting}
The \texttt{coupling\_values} dictionary maps coupling names to NumPy
arrays (e.g.\ $F_{abc} \to$ a rank-3 tensor), and
\texttt{fixed\_indices} pins observable component indices (e.g.\
\texttt{\{`a': 1, `b': 1\}} to select $\xi_{22}$).

For two-point correlators, the integration is driven by
\begin{lstlisting}
integrate_two_point_qmc(
    integrands: list[DiagramIntegrand],
    t_f: float,
    positions: dict[str, float],
    cache: PropagatorCache,
    t_min: float = 0.0,
    n_samples: int = 2**14,
    seed: int | None = None,
    external_times=None,
) -> tuple[float, float]
\end{lstlisting}
which returns the estimated value and standard error.
For up to five integration variables, tensor-product Gauss--Legendre
quadrature on the causally ordered simplex is recommended and is what the
worked examples use; the QMC backend, the default, is intended for
higher-dimensional integrals (sixth order and beyond).
The full list of keyword arguments is documented in the repository.

\section{Proof of the spatial multiplicity formula}
\label{app:proofs}

We prove \cref{eq:multiplicity}: for a given spatial topology (a fixed
set of $R$-edges and $C$-edges between spatial points), the number of
distinct Wick pairings that realise this topology is
\begin{equation}
  \mu = \frac{\displaystyle\prod_v m_v!}
             {\displaystyle 2^{N_{\mathrm{self}}} \cdot \prod_e k_e!}\,,
  \tag{\ref{eq:multiplicity}}
\end{equation}
where:
\begin{itemize}
  \item $m_v$ is the number of $\varphi$-operators at vertex $v$
    available for $C$-pairing (i.e., after the $R$-assignment has
    consumed some $\varphi$-operators);
  \item $N_{\mathrm{self}}$ is the total number of $C$-self-loops
    (edges with both endpoints at the same spatial point);
  \item $k_e$ is the multiplicity of each distinct $C$-edge (the
    number of parallel $C$-propagators between the same pair of
    spatial points).
\end{itemize}

\begin{proof}
The $R$-assignment is fixed: each $\psi$-operator at a given spatial
point is assigned to a specific $\varphi$-operator at another point.
Once the $R$-edges are determined, the remaining task is to pair the
leftover $\varphi$-operators via $C$-propagators in all ways consistent
with the prescribed spatial $C$-edge set.

At each vertex $v$, there are $m_v$ remaining $\varphi$-operators
(distinguishable by their unique UIDs). These operators must be
distributed among the $C$-edges incident to $v$: if edge $e$ connects
vertex $v$ to vertex $w$ with multiplicity $k_e$, then $k_e$ of the
$m_v$ operators at $v$ are assigned to edge $e$.

\paragraph{Numerator: operator permutations}
At vertex $v$, the $m_v$ operators can be arranged in $m_v!$ orderings.
Each ordering assigns operators to $C$-edge slots in sequence.
Across all vertices, the total number of arrangements is
$\prod_v m_v!$.

\paragraph{Denominator: overcounting from symmetries}

\emph{Self-loop symmetry.} For each $C$-self-loop at vertex $v$ (both
endpoints at the same point), swapping the two operators in the pair
produces the same contraction, since $C$ is symmetric:
$\vev{\varphi_a\,\varphi_b} = \vev{\varphi_b\,\varphi_a}$. Each
self-loop contributes a factor of $2$ to the overcounting, giving
$2^{N_{\mathrm{self}}}$ in total.

\emph{Parallel-edge symmetry.} For each pair of spatial points
$(v, w)$ connected by $k_e$ parallel $C$-edges, permuting the $k_e$
edges among themselves produces the same topology (since all edges
connect the same pair of points with the same propagator type). This
contributes a factor of $k_e!$ per edge group.

Dividing by both overcounting factors yields
\begin{equation}
  \mu = \frac{\prod_v m_v!}{2^{N_{\mathrm{self}}} \cdot \prod_e k_e!}\,.
\end{equation}

Observable operators (external points) are fixed and not permuted, so
they do not contribute to the multiplicity. Only operators at
interaction vertices enter the formula.
\end{proof}

\section{The \ito{} prescription: Jacobian of the delta functional and $Z_S = 1$}
\label{app:ZS-unity}

Both results of this appendix rest on one property of the \ito{}
prescription: a closed loop of retarded propagators vanishes. Around a
cycle of $k$ of them, $\RespOp(t_1,t_2)\cdots\RespOp(t_k,t_1)$, the
Heaviside factors impose $t_1 \geq t_2 \geq \cdots \geq t_k \geq t_1$, so
all times coincide and the product is proportional to $\Theta(0)^k$, which
vanishes for $\Theta(0) = 0$ (\cref{sec:ito}); a loop of length one is
$\RespOp(\bm{z},\bm{z}) \propto \Theta(0) = 0$ directly.

\subsection{Jacobian of the delta functional}
\label{app:jacobian}

\Cref{eq:delta-functional} uses the delta functional of the equation in
place of that of the solution. With
$f[\varphi] \equiv \partial_t\varphi - A\varphi - \mathcal{F}[\varphi]$ the
two are related by
$\delta[\varphi - \varphi_\eta] = J[\varphi]\,\delta[f[\varphi] - \eta]$,
where $J[\varphi] = \operatorname{Det}(\delta f/\delta\varphi)$ is the
Jacobian of the map $\varphi \mapsto f[\varphi]$. \Cref{eq:delta-functional}
therefore omits the factor $J[\varphi]$, and its normalisation is
$\int\mathcal{D}\varphi\,P[\varphi] = \langle 1/J[\varphi_\eta]\rangle_\eta$
rather than one. A field-independent $J$ changes only the overall constant,
which cancels in the normalised expectation value
\cref{eq: average O with S}; a field-dependent $J$ changes the relative
weight of configurations and cannot be absorbed.

Which case applies follows from
$\delta f/\delta\varphi = (\partial_t - A) - \mathcal{F}'
= (\partial_t - A)(1 - \RespOp\,\mathcal{F}')$, with
$\mathcal{F}' = \delta\mathcal{F}/\delta\varphi$ and
$\RespOp = (\partial_t - A)^{-1}$ the retarded Green's function of the
linear part (the response propagator of \cref{sec:propagators}), so that
$J = \operatorname{Det}(\partial_t - A)\,
\operatorname{Det}(1 - \RespOp\,\mathcal{F}')$. The first factor is
field-independent: it is the Jacobian of the linear theory, and
normalising the measure on that theory, as assumed for $Z_S = 1$ after
\cref{eq: average O with S}, absorbs it. The second has
$\ln\operatorname{Det}(1 - \RespOp\mathcal{F}')
= -\sum_{n\geq1}\operatorname{Tr}[(\RespOp\mathcal{F}')^n]/n$, and each
term is a closed loop of $n$ retarded propagators, which vanishes by the
loop property above: $J$ is field-independent. In the time-sliced integral
the Jacobian matrix is then lower-triangular with unit diagonal, so
$J = 1$ exactly, and \cref{eq:delta-functional} is normalised for any
driving-field statistics,
$\int\mathcal{D}\varphi\,P[\varphi] = \int\mathcal{D}\eta\,P[\eta] = 1$;
by \cref{eq: physical field prob 2} this is $Z_S = 1$. In the Stratonovich
(mid-point) discretisation $\Theta(0) = \tfrac12$, the one-loop term
survives as $\exp(-\tfrac12\operatorname{Tr}\mathcal{F}')$ and must be
kept in the action \cite{Hertz2017,ChowBuice2015}; the loops of length
two and more still vanish, their coincidence set having measure zero.

\subsection{$Z_S = 1$ diagram by diagram}
\label{app:ZS-proof}

Beyond the normalisation above, every vacuum diagram vanishes
individually: for local vertices, for the $\psi$-only non-local vertices
of \cref{eq:Sint-K} and for equal-time non-local vertices. This complements the path-integral argument of
\cref{app:jacobian} and of \cite{Martin1973,Janssen1976,DeDominicis1976}.

\begin{proof}
The zeroth-order term of $Z_S = \sum_n (-1)^n/n!\,\vevfree{\Sint^n}$
is unity. At order $n \geq 1$, every Wick contraction involves only
vertex fields (no external legs). Since
$\vevfree{\psi\,\psi} = 0$, each $\psi$ must contract with a
$\varphi$ via a retarded $R$-propagator, defining a directed
vertex-level multigraph $G_V$: an edge $\alpha \to \beta$ for each
$R$-contraction of $\varphi$ at vertex~$\alpha$ with $\psi$ at
vertex~$\beta$.

Each vertex contributes $q \geq 1$ response fields, so every node
has in-degree $\geq 1$. For \emph{local} vertices, the \ito{}
rule $R(\bm{z}, \bm{z}) = 0$ forbids self-loops. The \emph{non-local}
vertices of \cref{eq:Sint-K} carry response fields only and admit no
self-contraction; a non-local vertex with both field types at a common
time could self-contract, which is a directed cycle of length~1. In
either case, following incoming edges backwards from any node must
revisit a node within $n$ steps (pigeonhole), producing a directed
cycle.

Every directed $R$-cycle vanishes by the loop property above. Since every
vacuum diagram contains at least one $R$-cycle, $\vevfree{\Sint^n} = 0$
for all $n \geq 1$, hence $Z_S = 1$ order by order.
\end{proof}

Consequently, the perturbative expansion
$\vev{O}_S = \sum_{n=0}^{N} (-1)^n/n!\,\vevfree{O\,\Sint^n}$
requires no normalisation at any truncation order.

\section{Mean-field subtraction and initial conditions}
\label{app:mean-field}

\subsection{Mean-field subtraction}
\label{app:mean-field-subtraction}

The perturbative framework of \cref{sec:msr-action} assumes a
zero-mean driving field and treats the nonlinear terms $\mathcal{F}$ as
perturbations about a linear theory. In practice, several common
situations require a preliminary mean-field subtraction before applying
the formalism:
\begin{enumerate}
  \item A driving field with non-zero mean.
  \item A dynamics whose lowest-order nonlinearity is quadratic rather
    than linear (i.e., no linear term in $\mathcal{F}$).
  \item Non-trivial initial conditions.
\end{enumerate}
In such cases, one first solves the deterministic (noise-averaged)
problem with the given initial conditions, obtaining a background
solution $\bar\varphi$. Rewriting the Langevin equation
(\cref{eq:langevin}) in terms of the fluctuation
$\delta\varphi = \varphi - \bar\varphi$ yields an effective equation of
the same Langevin form, but now with a zero-mean driving field, a
linear operator given by the linearisation of the full dynamics about
$\bar\varphi$, renormalised nonlinear couplings, and trivially zero
initial conditions. After this subtraction, the perturbative formalism
of \cref{sec:msr-action} applies directly to the fluctuation field
without further complications.

\subsection{Role of initial conditions}
\label{app:initial-conditions}

The propagators $\CorrOp$ and $\RespOp$ defined in
\cref{sec:propagators} encode the dynamics generated by the driving
field from $t_{\min}$ onward; they carry no information about the
initial state of the system. More generally, the MSR path integral
computes expectation values conditioned on (or averaged over) a
prescribed initial distribution at $t = t_{\min}$, and any
initial-state correlations enter as a separate contribution. For the
two-point function, for example, the full correlator is
\begin{multline}
  \label{eq:initial-condition}
  \vev{\varphi_a(t_1)\,\varphi_b(t_2)} =
  \bigl[\RespOp(t_1, t_{\min})\,C_0\,\RespOp^T(t_2, t_{\min})\bigr]_{ab} \\
  + \CorrOp_{ab}(t_1, t_2)\,,
\end{multline}
where $C_0$ is the initial correlation matrix and spatial arguments are
suppressed for brevity. Higher-order moments receive analogous
initial-state contributions.

When the linear dynamics is stable (all eigenvalues of $\mathbf{A}$
have negative real part) and $t_{\min} \to -\infty$, the initial-state
terms decay and the MSR propagators alone yield steady-state
expectation values. For finite-time problems (such as the weak lensing
application of \cite{Zhang2025_SFTLensing}), initial conditions may be
significant and should be included separately, either as explicit
boundary terms or by augmenting the action with an initial-state
vertex.

In practice, however, the mean-field subtraction described in
\cref{app:mean-field-subtraction} often removes this complication
entirely: once the background solution absorbs the initial conditions,
the fluctuation field starts from zero and the MSR propagators capture
the full correlator without additional boundary terms.

\section{Direct numerical simulation}
\label{app:simulation}

This appendix describes the Heun simulation used to validate the
perturbative results in \cref{sec:examples}; the simulations of
\cref{sec:example2} are described at its end.

\subsection{Noise generation}

The key technical step is generating the temporally and spatially
correlated driving field $\eta_a(x,t)$. Exploiting the separable
kernel $\cumulant_{ab}(x,t;x',t') = \delta_{ab}\,K_t(t,t')\,K_x(x,x')$
with $K_t(t,t') = \lambda\,e^{-|t-t'|/\sigma_t}$ and
$K_x(x,x') = e^{-|x-x'|/\sigma_x}$, the spatial correlations are
generated via Cholesky factorisation of $K_x$, while the temporal
correlations are produced exactly by an autoregressive recursion
that exploits the Markov property of the exponential kernel:
\begin{multline}
  \eta_a(x_i,t_{k+1})
  = \rho\,\eta_a(x_i,t_k)
  + \sigma_{\mathrm{innov}}\!\sum_j L^{(x)}_{ij}\,\varepsilon_{a,j,k}\,, \\
  \varepsilon_{a,j,k}\sim\mathcal{N}(0,1)\,,
  \label{eq:ar1-noise}
\end{multline}
where $\rho = e^{-\Delta t/\sigma_t}$,
$\sigma_{\mathrm{innov}} = \sqrt{\lambda(1-\rho^2)}$, and
$K_x = L^{(x)}(L^{(x)})^{\!\top}$.
The initial condition is drawn from the stationary distribution,
$\eta_a(x_i,t_0) = \sqrt{\lambda}\sum_j L^{(x)}_{ij}\,z_{a,j}$.
This $\mathrm{AR}(1)$ scheme reduces the noise generation cost from
$O(n_{\mathrm{steps}}^2)$ (full temporal Cholesky) to $O(1)$ per
time step.

\subsection{Time stepping}

The fields are evolved using Heun's method (a predictor--corrector
scheme with second-order weak convergence):
\begin{align}
  \tilde\varphi_a^{k+1} &= \varphi_a^k + \Delta t\,
    f_a(\varphi^k, \eta^k)\,,
    \label{eq:heun-predictor}\\
  \varphi_a^{k+1} &= \varphi_a^k + \tfrac{\Delta t}{2}\,
    \bigl[f_a(\varphi^k, \eta^k)
    + f_a(\tilde\varphi^{k+1}, \eta^{k+1})\bigr]\,,
    \label{eq:heun-corrector}
\end{align}
where $f_a(\varphi,\eta) = A_{ab}\,\varphi_b
+ F^{(3)}_{abc}\,\varphi_b\varphi_c + \eta_a$ is the full
right-hand side and the spatial index is suppressed. Zero initial
conditions $\varphi_a(x,0) = 0$ are used, with step size
$\Delta t = 0.02$.

\subsection{Statistical estimation}

The two-point correlator
$\xi_{ab}(r,t) = \langle\varphi_a(0,t)\,\varphi_b(r,t)\rangle$ is
estimated from the ensemble average over $N$ independent realisations.
For Gaussian fields the Monte Carlo standard error is bounded by the
Isserlis (Wick) estimate
\begin{equation}
  \sigma_\xi(r) = \frac{\sqrt{C_0^2 + \xi(r)^2}}{\sqrt{N}}\,,
  \label{eq:wick-noise}
\end{equation}
where $C_0 = C_{aa}(0,t;0,t)$ is the equal-point variance. For
$N = 100\,000$ realisations this gives
$\sigma_\xi/\xi \lesssim 0.5\%$ at $r = 0$.

\subsection{Simulations of the shot-noise example}
\label{app:sim-demo3}

The shot-noise example of \cref{sec:example2} uses three simulations, all
with the driving field sampled exactly from its Poisson events (the
truncated event window contributes at most $7.6\times10^{-11}$ at
$F = 0$). At $F = 0$ the response to each exponential pulse is analytic,
so the physical field is evaluated per event without time stepping
($1.2\times10^{6}$ realisations). At finite $F$ the interaction is
integrated on top of the exactly sampled free field with a second-order
exponential time-differencing scheme (ETDRK2 \cite{CoxMatthews2002}),
checked by halving the step from $\Delta t = 0.01$ to $0.005$;
$\xi_{12}(t)$ uses six seeds of $4\times10^{5}$ realisations, $\xi_{12}(r)$
one seed of $10^{6}$ and $\xi_{11}$ one seed of $4\times10^{5}$. The
realisation of \cref{fig:ex2-waterfall} is an illustration integrated with
the Heun scheme at $\Delta t = 0.004$ and plays no part in the comparisons.


\section*{CRediT authorship contribution statement}
\textbf{Zheng Zhang:} Conceptualization, Methodology, Software,
Validation, Visualization, Writing, Review
\& Editing.

\section*{Data availability}
\sftwick{} version~0.4.2, the version described in this paper, is
available from PyPI (\texttt{pip install sft-wick==0.4.2}). The source
code, the worked examples with their configurations, simulation scripts
and cached simulation data (\texttt{examples/demo1},
\texttt{examples/demo2}, \texttt{examples/demo3},
\texttt{examples/paper\_assets}) and the
validation catalogue are openly available under the BSD-3-Clause
licence at \url{https://github.com/StatFieldTheory/sft-wick} and are
permanently archived on Zenodo~\cite{sftwick_software}
(version~0.4.2:~\href{https://doi.org/10.5281/zenodo.22284796}{10.5281/zenodo.22284796};
concept DOI:~\href{https://doi.org/10.5281/zenodo.20776358}{10.5281/zenodo.20776358},
which resolves to the latest archived release).


\section*{Declaration of generative AI and AI-assisted technologies in the manuscript preparation process}
During the preparation of this work the author used Claude Code
(Anthropic) in two ways: in the manuscript, to identify issues with
grammar and narrative flow and to improve the figures; and in the
development of the \sftwick{} software, as a coding assistant that
drafted parts of the source code, in-code commentary, tests and
documentation under the author's direction. The formalism, the
algorithms, the design of the validation suite and the interpretation
of all results are the author's. After using this tool, the author
reviewed and edited the content as needed; all code, whether
AI-assisted or not, is validated by the test suite catalogued in
\cref{app:validation}, and the author takes full responsibility for the
content of the publication and of the software.


\makeatletter
\newcommand\arxivrightcolumnref{17}
\newcount\arxiv@refcount
\AddToHook{env/thebibliography/begin}{%
  \let\arxiv@bibitem\bibitem
  \def\bibitem{%
    \global\advance\arxiv@refcount\@ne
    \ifnum\arxiv@refcount=\arxivrightcolumnref\relax
      \vfill\eject
    \fi
    \arxiv@bibitem}}
\makeatother
\bibliographystyle{elsarticle-num}
\bibliography{bibliography_r1}

\end{document}